\documentclass[12pt,preprint]{emulateapj}

\newcommand{\ts}{\thinspace}
\newcommand{\simless}{\mathbin{\lower 3pt\hbox
     {$\rlap{\raise 5pt\hbox{$\char'074$}}\mathchar"7218$}}}
\newcommand{\simgreat}{\mathbin{\lower 3pt\hbox
     {$\rlap{\raise 5pt\hbox{$\char'076$}}\mathchar"7218$}}}

\newcommand{\about}    {$\sim$\ts}
\newcommand{\aboutless}{$\simless$\ts}
\newcommand{\aboutmore}{$\simgreat$\ts}

\newcommand{\JK}{$J-K_{\rm s}$}
\newcommand{\JKO}{($J-K_{\rm s}$)$_{\rm o}$}

\newcommand{\RM}{$R_{24/4.5}$}

\newcommand{\IRAS}{{\it IRAS}}
\newcommand{\ISO}{{\it ISO}}
\newcommand{\Spitzer}{{\it Spitzer}}
\newcommand{\msun}{\ts M$_\odot$}

\newcommand{\etal}{\ts et~al.}

\shortauthors{Carpenter \etal}
\shorttitle{Debris Disks in Upper Sco}

\begin{document}

\title{Debris Disks in the Upper Scorpius OB Association}

\author{John M. Carpenter\altaffilmark{1}}
\author{Eric E. Mamajek\altaffilmark{2}}
\author{Lynne A. Hillenbrand\altaffilmark{1}}
\author{Michael R. Meyer\altaffilmark{3}}

\altaffiltext{1}{California Institute of Technology, Department of Astronomy, MC 105-24, Pasadena, CA 91125}
\altaffiltext{2}{University of Rochester, Department of Physics and Astronomy, Rochester, NY, 14627-0171}
\altaffiltext{3}{Institute for Astronomy, ETH, CH-8093 Zurich, Switzerland}

\begin{abstract}

We present MIPS 24\micron\ and 70\micron\ photometry for 205 members of the
Upper Scorpius OB Association. These data are combined with published MIPS
photometry for 15 additional association members to assess the frequency of
circumstellar disks around 5 Myr old stars with spectral types between B0 and
M5. Twelve stars have a detectable 70\micron\ excess, each of which also has a
detectable 24\micron\ excess. A total of 54 stars are identified with a
24\micron\ excess more than 32\% above the stellar photosphere. The MIPS
observations reveal 19 excess sources -- 8 A/F/G stars and 11 K/M stars -- that
were not previously identified with an 8\micron\ or 16\micron\ excess. The lack
of short-wavelength emission and the weak 24\micron\ excess suggests that these
sources are debris systems or the remnants of optically thick primordial disks
with inner holes. Despite the wide range of luminosities of the stars hosting
apparent debris systems, the excess characteristics are consistent with all
stars having dust at similar orbital radii after factoring in variations in the
radiation blowout particle size with spectral type. The results for Upper Sco
are compared to similar photometric surveys from the literature to re-evaluate
the evolution of debris emission. After considering the completeness limits of
published surveys and the effects of stellar evolution on the debris
luminosity, we find that the magnitude of the 24\micron\ excess around F-type
stars increases between ages of 5 and 17~Myr as found by previous studies, but
at \aboutless 2.6$\sigma$ confidence. For B7-A9 and G0-K5 stars, any variations
in the observed 24\micron\ excess emission over this age range are significant
at less than 2$\sigma$ confidence.

\end{abstract}

\maketitle

\section{Introduction}

Dusty debris observed around main sequence stars is a potential diagnostic
of planetary systems. The current paradigm is that the debris originates from
the collisional grinding of planetesimals to micron-sized grains, where the
collision rate is dictated by gravitational interactions between planets and a
planetesimal belt \citep{Williams94}. If this hypothesis is correct, the
spatial asymmetries in the debris that have been revealed by high resolution
imaging \citep[e.g.][]{Holland98} encodes information on the eccentricity, 
mass, and even migration of the orbiting planets \citep{Liou99,Wyatt03}.
Unfortunately the resolution and sensitivity required to image the debris can
be achieved for only a few disks with current instruments.

A complementary approach to detailed studies of individual disks is to study
the ensemble properties of debris systems. Debris disks have been studied
extensively over the entire sky using \IRAS\
\citep{Backman93,Lagrange00,Rhee07} and in targeted regions using \ISO\
\citep{Habing01,Spangler01,Dominik03}. The \Spitzer\ Space Telescope
\citep{Werner04} has expanded on these studies to produce a vast database of
debris disks \citep[e.g.][see also reviews by \citealt{Werner06},
\citealt{Meyer07} and
\citealt{Wyatt08}]{Rieke05,Bryden06,Su06,Gautier07,Currie08a,Carpenter09} that
encompass a broad range of spectral types (B-M stars), environments (clusters,
associations, field stars), and ages (3~Myr to 10~Gyr). Comparison of these
data to theoretical models have yielded insights on the planetesimal belts that
produce the debris dust \citep{Wyatt07b}, the collisional history of
planetesimal belts \citep{Dominik03,Wyatt07a,Lohne08} and the formation of
planetary systems \citep[and references therein]{Kenyon08}.

A critical issue emerging from \Spitzer\ studies is establishing when the
debris phenomenon is initiated. In the Kenyon \& Bromley models, a planetesimal
belt produces low levels of debris emission in the early stage of
planetary accretion. At a given orbital radius, dust production reaches a
maximum when 1000-3000~km sized bodies form and ignite the destructive
collisional cascade \citep{Kenyon04}. As planet formation propagates
through the disk, the disk is depleted of planetesimals through repeated 
collisions and the debris production eventually declines. In theory, the
evolution of the debris emission can constrain the formation time of planets
and the collisional cascade. Indeed, based on \Spitzer\ observations,
\citet{Hernandez06} and \citet{Currie08a} have suggested that the debris
luminosity peaks around 10-30~Myr for A-F stars before declining toward older 
ages \citep{Rieke05}.

Additional observations are needed to validate previous conclusions since few
surveys for debris emission at ages of \aboutless 10~Myr have been completed
\citep{Hernandez06,Carpenter06,Gautier08,Currie09a}. The Upper Scorpius OB
Association (hereafter, Upper Sco) is an important region in studies of disk
evolution since it provides a snapshot of disk properties at an age of \about
5~Myr when most optically thick disks have dissipated \citep{Hernandez07a}.
\Spitzer\ observations can then probe the onset of debris production as planets
form. Since association members have been identified from B to M stars, disk
properties can be investigated over nearly two orders of magnitude in stellar
mass. Finally, at a mean distance of 145~pc \citep{deZeeuw99}, Upper Sco is
more than a factor of two closer than other clusters or associations of similar
age. These traits permit sensitive photometric studies of debris properties.

In previous studies, we investigated the 4.5-16\micron\ photometric
\citep[hereafter Paper I]{Carpenter06} and 5-35\micron\ spectroscopic
\citep{Dahm09} properties of circumstellar disks in Upper Sco. The principle
result from these studies is that the circumstellar disks detected at
wavelengths shortward of 16\micron\ show a dichotomy with spectral type.
Circumstellar disks around late-type stars (K and M spectral types) are likely
optically thick primordial disks formed as a result of the star formation
process, while early-type stars (B and A types) appear to be surrounded by
debris disks. Surprisingly, no disks were detected around the F and G stars.

The photometric data presented in Paper~I spanned wavelengths between
4.5\micron\ and 16\micron. To probe for cooler dust, we present here 24\micron\
and 70\micron\ photometric observations of 205 Upper Sco members. These new
\Spitzer\ observations of Upper Sco are presented in Section~\ref{obs}. In
Section~\ref{excess}, we use these data to identify stars that have excesses
in the 24\micron\ and 70\micron\ bands. The nature of the excess sources are
investigated in Section~\ref{nature}, and properties of debris systems
around stars with a wide range in mass are compared in Section~\ref{prop}. In
Section~\ref{evolution}, we compare the Upper Sco results with published
\Spitzer\ surveys to investigate the evolution of debris systems.

\section{Observations and Data Reduction}
\label{obs}

The selection criteria for the Upper Sco sample are described in Paper~I.
Briefly, the sample was constructed from membership lists established by
Hipparcos astrometry \citep{deZeeuw99}, color-magnitude diagrams
\citep{Preibisch99,Preibisch02}, and X-ray surveys
\citep{Walter94,Martin98,Preibisch98,Kunkel99,Kohler00}. Since membership
selection was based upon stellar properties (proper motions, photospheric
colors, X-ray activity) unrelated to circumstellar disks, the sample is
believed to be unbiased with respect to the presence or absence of disks. From
the compiled list, we selected 205 stars\footnote{\citet{Bouy09} found that
the star [PBB2002] USco 161021.5$-$194132 deviates from the mean proper motion
for Upper Sco and may be an interloper. The reduced $\chi^2$ from the 2MASS
point-spread-function fitting photometry is the second highest among the
nearest 100 stars to this source that have a 2MASS quality flag of AAA. The
astrometry may therefore be contaminated by a binary. The luminosity of this
star is also higher than the 10 nearest stars of the same spectral type
\citep{Preibisch02}, consistent with a binary contaminant. Given that the
lithium equivalent width \citep{Preibisch02} and apparent brightness are
consistent with Upper Sco membership, we retain the source in our sample.} to
sample spectral types between B0 and M5. This sample was supplemented with 15
Upper Sco members with G/K spectral types that were observed by the FEPS
(Formation and Evolution of Planetary Systems) \Spitzer\ Legacy program
\citep{Meyer06}. The FEPS targets were selected based on similar criteria as
our study and should also be an unbiased sample with respect to the presence or
absence of disks.

Other Upper Sco members have been observed with \Spitzer\ but are not included
in our sample. The Herbig Ae/Be star HIP~79476 \citep{Hernandez05} was observed
with the \Spitzer\ Infrared Spectrograph (IRS), but not with MIPS. We did not
include the G6/G8 star HD~143006, which was also observed by the FEPS program
as part of survey of gas in disks. HD~143006 was recognized as a Upper Sco
member based on an \IRAS\ excess \citep{Odenwald86} and thus would bias the
sample. We also do not use the three B stars observed by \citet{Rieke05} and
the five F/G stars observed by \citet{Chen05} since IRAC data are not available
for these sources, which is needed for the analysis presented in
Section~\ref{excess}. Finally, we did not include brown-dwarfs in Upper Sco
\citep{Scholz07}.

In Paper~I, IRAC (4.5\micron\ and 8\micron) and IRS peak-up (16\micron)
photometry was presented for 218 of the 220 stars. IRAC photometry for one
additional Upper Sco member observed by FEPS (ScoPMS~52) is now available
\citep{Carpenter08}. The remaining star (HIP~80112) was not observed with IRAC
since it would have saturated the detector. The 4.5\micron\ flux density
for this source was estimated from ground-based $M$-band observations
\citep{vanderBliek96}.

We present here MIPS 24\micron\ and 70\micron\ photometry for 205 stars in
Upper Sco. The data reduction procedures follow that used by the FEPS program
as described in \citet{Carpenter08}, and a only summary of the salient steps is
provided here. Table~\ref{tbl:phot} presents the 24\micron\ and 70\micron\ MIPS
photometry for the 205 stars observed for this study. MIPS photometry for the
remaining 15 stars are listed in \citet{Carpenter08}. 

\subsection{MIPS 24\micron}
\label{obs:mips24}

MIPS~24\micron\ observations for 205 stars were obtained in photometry mode.
The exposure time per image (either 3 or 10~sec) and the number of dithered
images were set to achieve a signal-to-noise ratio of at least 10 on the
expected photospheric brightness. MIPS 24\micron\ images were processed
starting with the Basic Calibrated Data (BCD) products produced by the
\Spitzer\ Science Center (SSC) pipeline version S17. Individual BCD images were
combined to derive a flat field that removed long-term gain changes in the MIPS
array. Flat field images were derived from stars that are not surrounded by
nebulosity. If nebulosity is present, a flat-field image from another star was
used that was observed close in time and with the same exposure time.

Point(source)-response-function (PRF) photometry was performed with the MOPEX
package \citep{Makovoz05}. The empirical PRF distributed with MOPEX was fitted
to the individual BCD images simultaneously (as opposed to the mosaicked image)
using a fitting area of 21$\times$21 pixels (1 pixel $\approx$ 2.55\arcsec) for
most sources. A 5$\times$5 pixel fitting area was used for sources that have
spatially variable nebulosity near the point source position. From visual
inspection of the mosaicked images, the PRF of the Upper Sco target overlaps
occasionally with that from a nearby source. These contaminating sources were
fitted with a PRF simultaneously. The measured 24\micron\ flux densities and
internal uncertainties from the PRF fit are presented in Table~\ref{tbl:phot}.
The images were processed with a calibration factor of 0.0447~MJy~sr$^{-1}$.
Following \citet{Engelbracht07}, we adopt a calibration uncertainty of 4\%. 

Cirrus and extragalactic sources may contaminate the MIPS photometry and create
the appearance of an infrared excess. Since we expect the emission from a
circumstellar disk to be point-like and centered on the star at the distance of
Upper Sco, contaminated 24\micron\ photometry can be identified from emission
that is extended or offset from the stellar position.

The maximum astrometric offset observed between 2MASS and the \Spitzer\
24\micron\ position is 1.6\arcsec\ for HIP~78820. The dispersion in the offsets
is the same (\about 0.4 to 0.55\arcsec) for stars with and without infrared
excesses (see Section~\ref{excess:mips24}), indicating that any contamination
is not systematically producing offsets in the measured coordinates. This
empirical result is consistent with the contamination anticipated from
extragalactic source counts \citep{Papovich04}. For each star in the sample, we
computed the expected number of galaxies located within the
full-width-at-half-maximum (FWHM) size (6\arcsec) of the MIPS~24\micron\ PRF
that can produce a 30\% photometric excess at 24\micron. (As described in
Section~\ref{excess:mips24}, 30\% is approximately the minimum excess
detectable with these observations.) We find that \about 1 star may be
contaminated by an extragalactic source bright enough to produce such an
excess.

A more significant level of contamination may arise from cirrus and nearby
stars. We used both quantitative techniques and visual inspection of the images
to identify such cases. First, we compared PRF photometry to aperture
photometry obtained with a 6~pixel (1~pixel = 2.55\arcsec) aperture diameter,
which is approximately 2.5 times the FWHM size of the MIPS 24\micron\ PRF. For
38 out of the 220 stars in the sample, the aperture photometry deviated from
the PRF photometry by more than 10\%. We chose 10\% as the threshold since such
deviations would impact the significance the 24\micron\ excesses identified in
Section~\ref{excess:mips24}. Visual inspection of the images revealed that in
the vast majority of cases, the difference between the aperture and PRF
photometry is most likely due to low signal-to-noise since the aperture size is
much larger than the optimal extraction size for relatively faint sources
\citep{Naylor98}. The photometry was deemed questionable for 6 stars
based on the presence of extended nebulosity (HIP 77840, HIP 78265, HIP 79404,
and $[$PBB2002$]$ USco J161052.4$-$193734) or because the 24\micron\ emission,
while centered on the stellar position, is extended (HIP 77858 and HIP 80338).
In addition, for the star ScoPMS 17, a known contaminating source is resolved
in the 16\micron\ images (see Paper~I) but is unresolved at 24\micron. These
seven stars are flagged in Table~\ref{tbl:phot}.

\subsection{MIPS 70\micron}
\label{obs:mips70}

MIPS~70\micron\ observations were obtained in photometry mode with an exposure 
time of 10~sec and the small field size dither pattern. The number of cycles 
was fixed at 4 for all stars. MIPS~70\micron\ images were processed with 
SSC pipeline version S17 that removes the bias,
subtracts a dark image, applies a flat field correction, and linearizes the
pixel response. Individual BCD images were mosaicked with the Germanium
Reprocessing Tools (GeRT) software package S14.0 version 1.1 developed at the
SSC. A $40''\times40''$ region centered on the source position was excluded
when computing the column and time filtering such that the filtering process is
not biased by the presence of a bright source. Filtered images were formed into
mosaics with MOPEX \citep{Makovoz05}. 

Aperture photometry was performed on the MIPS~70\micron\ mosaics with a custom
version of IDLPHOT using an aperture radius of 16\arcsec\ (4 pixels on the
coadded images). The sky-level was computed as the mean pixel value in a sky
annulus that extended from 40$''$ to 60$''$. The aperture was centered on the
expected stellar position computed from the world coordinate system keywords
contained in the FITS image headers. No centroiding was performed since the
signal-to-noise ratio of most 70\micron\ measurements are less than 3. Visual
inspection of the 70\micron\ mosaics identified 19 images where a point source
was located within the outer sky annulus or the aperture radius, but offset
from the 2MASS stellar position by more than 4\arcsec. Before measuring
the aperture photometry, a PRF was fitted to the apparent contaminating source 
and subtracted from the image using MOPEX. 

The MIPS~70\micron\ photometry and internal uncertainties are presented in 
Table~\ref{tbl:phot}. Stars for which the 70\micron\ photometry was measured on
PRF-subtracted images are marked in the table. The adopted calibration factor
is 702.0~MJy sr$^{-1}$ / (DN s$^{-1}$) with an uncertainty of 7\% as reported
on the SSC MIPS calibration web
pages.\footnote{http://ssc.spitzer.caltech.edu/mips/calib}

Of the 220 stars in the sample, 15 have a 70\micron\ signal-to-noise ratio $\ge
3$. Photometry for two of these stars is compromised. ScoPMS~17 is contaminated
by a nearby source resolved in the 16\micron\ images (see Paper~I). HIP~80338
is surrounded by nebulosity and it is ambiguous if the 70\micron\ emission is
associated with the star. In the remainder of this paper, we do not include
these two stars in analysis of the 70\micron\ data.

\section{Sources with Infrared Excesses}
\label{excess}

We now combine the 24\micron\ and 70\micron\ photometry for 220 Upper Sco
members (i.e. 205 stars from this study, and 15 stars from FEPS; see
Section~\ref{obs}) with 2MASS $J$ and $K_{\rm s}$ \citep{Skrutskie06} and IRAC
4.5\micron\ \citep[Paper~I,][]{Carpenter08} photometry to identify stars with
infrared excesses. All photometry was dereddened using the extinction law from
\citet{Cardelli89} assuming $R_V=3.1$ for $\lambda \le 2$\micron, the
\citet{Chapman09} reddening law for IRAC and MIPS~24\micron, and the reddening
law compiled by \citet{Mathis90} for 70\micron. The visual extinction toward
individual stars were obtained from the literature, or derived by us from
published spectral types and optical/2MASS photometry following the general
procedure described in \citet{Carpenter08}. The visual extinction is
less than 2~mag for 97\% of the stars, which implies a correction of less than
4\% when dereddening the 24\micron\ photometry. The adopted spectral types and
visual extinction for each star are listed in Table~\ref{tbl:excess}.

\subsection{24\micron\ excesses}
\label{excess:mips24}

Figure~\ref{fig:ccd} presents the 24\micron\ to 4.5\micron\ flux ratio
($\equiv$ \RM) as a function of dereddened \JK\ color for the Upper Sco sample.
The \JK\ color should represent the stellar photosphere for most sources since
only two sources have an apparent $K_s$-band excess (Paper~I). We chose to
normalize the 24\micron\ photometry to the 4.5\micron\ band as a compromise
between reducing the uncertainties in extinction corrections and the range of
intrinsic photospheric colors (favoring longer wavelengths), and biasing the
results due to weak
emission from circumstellar dust (favoring shorter wavelengths). The locus of
points with log \RM\ $\approx-1.4$ have flux ratios roughly consistent with the
stellar photosphere for emission in the Rayleigh-Jeans limits. In practice, the
value of \RM\ varies systematically with \JKO, indicating changes in the
photospheric color with spectral type. A number of sources have values of \RM\
that lie substantially above this locus. These sources include the 35 stars
from Paper~I that have an 8\micron\ and/or 16\micron\ excess as indicated by
the black circles and crosses. 

We adopted an iterative procedure to determine the locus of photospheric 
colors in the \RM\ vs. \JK\ diagram. The trend between log(\RM) and \JKO\ color
was fitted using robust linear regression after first excluding the known 35
stars with 8\micron\ or 16\micron\ excesses and the 7 stars flagged with suspect
24\micron\ photometry. Sources that deviated from the best fit line by more 
than three times the dispersion of the residuals were removed, and the fit
was repeated until no additional sources were excluded. 

The best-fit relation is shown as the solid line in Figure~\ref{fig:ccd}. The 
dispersion about the best-fit line is $\sigma=0.03$~dex (i.e. 7\% of the 
photosphere) for the sources included in 
the fit. The largest deviation below the best-fit line is $-2.5\sigma$. Given 
the sample size of 220 stars, we expected only one source more discrepant than 
$-2.5\sigma$ for gaussian noise. The dispersion of 7\% is larger than the
median internal uncertainty (\about 2\%), suggesting that either the 
photometric precision is dominated by uncertainties not been quantified in our 
data reduction or that many of sources contain weak excesses. We conservatively
adopt a 7\% uncertainty on the \RM\ ratio for all sources. 

To identify stars that have a 24\micron\ excess, we required that the residuals
exceed the photospheric color by 4$\sigma$ = 0.12~dex, or a 32\% excess above
the photosphere. We adopted 4$\sigma$ as the cutoff since the dispersion in the
residuals can vary by \about 0.003~dex depending on which sample of stars are
used in the fit. By choosing a 4$\sigma$ cutoff, weak excesses are likely
significant even if the dispersion is underestimated by 0.003~dex. In total 54
stars have a value of \RM\ that exceeds the stellar photosphere by more than
4$\sigma$ and have reliable photometry. Of these 54 stars, 35 have an 8\micron\
and/or 16\micron\ excess, while 19 have a 24\micron\ excess only. These 19
sources are indicated by the filled stars in Figure~\ref{fig:ccd}. The sources
with excesses only at 24\micron\ span the full range of spectral types and
include 4 A/B-type stars, 1 F2 star, 3 G-type stars, 2 K-type stars, and 9
M-type stars.

In addition to the checks described in Section~\ref{obs:mips24}, we visually
re-inspected the 24\micron\ mosaics and the PRF subtracted images for the 54
stars with an infrared excess. Cutouts of the 24\micron\ images and
PRF-subtracted images for these sources are presented in
Figure~\ref{fig:images}. We identified 7 stars where nebulosity or source
confusion may compromise the photometry and create an apparent excess. Three
stars (HIP~77545, HIP~80024, and ScoPMS~45) are located in regions of extended
nebulosity, although the observed PRF is point-like. Four other stars
([PBB2002] USco J155729.9$-$225843, [PBB2002] USco J160532.1$-$193315,
[PBB2002] USco J160708.7$-$192733, and [PBB2002] USco J160900.0$-$190836) have
a neighboring source within \about 20\arcsec. We have re-reduced the data for
these sources using smaller sky annuli for the PRF fit, and examined various
aperture measures. In each case, the infrared excess is robust to these
variations in the data reduction. Also, in 4 of the 7 sources, the excess is
confirmed by IRAC 8\micron\ or IRS 16\micron\ photometry, where confusion and
nebulosity are less of a concern. We therefore classify these 7 stars as having
an 24\micron\ excess, but we note that the excesses around HIP~77545, [PBB2002]
USco J160708.7$-$192733, and ScoPMS~45 are detected only at 24\micron\ and are
otherwise unconfirmed.

The ratio of the observed 24\micron\ flux density to the photospheric value was
estimated from the observed photometry and the expected stellar colors. If a
source does not have an excess in the 4.5\micron\ or 8\micron\ bands (see
Paper~I), the observed-to-photospheric 24\micron\ ratio was computed from the
observed \RM\ ratio and the photospheric value determined from the \RM\ vs.
$(J-K_{\rm s})_{\rm o}$ relation shown in Figure~\ref{fig:ccd}. Otherwise, the
24\micron\ photospheric flux density was bootstrapped from the $K_{\rm s}$-band
photometry (or $J$-band if a $K_{\rm s}$ excess is present) using empirical
relations between spectral type and $J-K_{\rm s}$ color
\citep[Paper~I,][]{Dahm09}, spectral type and $K-M$ color\footnote{We compared
the empirical $K-M$ colors versus the model $K-[4.5]$ colors from the Stellar
Performance Estimation Tool (STAR-PET) available on the \Spitzer\ web site.
Between spectral types of A0 and G5, the maximum difference between the two
colors is 0.03~mag. For M0 stars, the color difference is 0.15~mag. A
comparison with the observational data indicates that the $K-M$ colors provide
a better match to the data.} compiled by \citet{Kenyon95}, and the \RM\ vs.
$(J-K_{\rm s})_{\rm o}$ relation. Table~\ref{tbl:excess} lists the ratio of the
observed-to-photospheric 24\micron\ flux densities for each star. 

\subsection{70\micron\ excesses}
\label{excess:mips70}

The photospheric flux density at 70\micron\ was estimated from the
24\micron\ flux density by assuming that the stellar emission is in the
Rayleigh-Jeans limit (i.e. $S_\nu \propto \nu^2$). If the source has a known
24\micron\ excess, the photospheric 24\micron\ flux density was estimated 
as described in Section~\ref{excess:mips24}. Otherwise, the observed 24\micron\
flux density was used.

Figure~\ref{fig:mips70_excess} presents a histogram of the measured
signal-to-noise ratio for the observed 70\micron\ infrared excess above the
stellar photosphere. The detected 70\micron\ flux density from the B0.5 star
HIP~78820 is consistent with the photospheric emission. Excluding the two stars
where the measured 70\micron\ flux density is suspect (see
Section~\ref{obs:mips70}), 12 stars have an apparent 70\micron\ excess greater
than 3$\sigma$. Sources with 70\micron\ excesses include 4 A/B stars, 6 K
stars, and 2 M stars. All 12 of these stars also have a 24\micron\ excess, as
well as an 8\micron\ or 16\micron\ excess identified in Paper~I.

\section{Nature of the Disk Candidates}
\label{nature}

The circumstellar dust producing the infrared excesses observed in Upper Sco
may originate from either a ``primordial'' or ``debris'' disk. Conceptually,
primordial disks contain copious amounts of gas and dust that are remnants of
the star formation process, while debris disks are gas-poor systems created by
the collisional shattering of planetesimals. 

Distinguishing primordial from debris dust is difficult observationally at an
age of 5~Myr. The inner portion of most primordial disks are optically thin by
this age as traced by 3-10\micron\ observations \citep{Mamajek04,Hernandez07a},
but the state of the outer disk as traced by wavelengths longer than 24\micron\
remain poorly characterized. The classic argument to support the debris
interpretation is that radiative and collisional processes can deplete the dust
on timescales much shorter than the stellar age assuming the disk is gas poor
\citep{Backman93}. Therefore, the dust must be replenished continuously,
presumably from a collisional cascade, to explain the observed infrared
excesses. This argument assumes that gas drag is negligible, which can
circularize the orbits of the dust grains and reduce the frequency of
destructive collisions. Models by \citet{Takeuchi01} indicate that
10~M$_\earth$ of gas distributed radially over \about 100~AU can prolong the
lifetime of grains around A-type stars to a couple of million years, which is
comparable to the age of Upper~Sco.

Unfortunately, observations of the gas component are not available for most of
the Upper Sco sample. Several stars in Upper Sco have strong H$\alpha$ emission
lines that suggest gas accretion is present, although most stars do not appear
to be accreting detectable levels of material \citep{Dahm09}. Yet even these
observations only trace gas in the inner disk and are not direct diagnostics 
of the total gas surface density. Extensive surveys for gas in 5-100~Myr stars
have been presented in other studies that are designed to trace the gas over a
range of orbital radii \citep{Zuckerman95,Najita05,Pascucci06}. Even these
observations, though, cannot rule out that gas drag has a significant influence
on the dust dynamics.

Given the lack of observations of the gas component, we rely on distinctions in
the spectral energy distributions between stars in Upper Sco and other known
sources to identify potential debris systems. In Paper~I, we compared the
spectral energy distributions of the Upper Sco sample between 2 and 16\micron\ 
with Herbig Ae/Be stars and classical T Tauri stars. We suggested that the
infrared excesses around the B/A stars represent debris systems, while the K/M
stars with 8\micron\ and 16\micron\ excesses originate from optically thick
disks, albeit with suppressed levels of mid-infrared emission relative to a
typical classical T Tauri star in Taurus. The 24\micron\ photometry and IRS
spectra for a subset of these sources support these conclusions \citep{Dahm09}. 

With the discovery of 19 additional disks in Upper Sco from the 24\micron\ 
photometric survey, we further investigate the range of disk characteristics.
Figure~\ref{fig:ic348} plots the 24\micron\ to
4.5\micron\ flux ratio versus the 8\micron\ to 4.5\micron\ flux ratio for stars
in Upper Sco with (solid black circles) and without (gray circles) 24\micron\
excesses. Qualitatively, a dichotomy is present in the excess characteristics.
One population, present only for stars with spectral types later than K0, has
both 8\micron\ and 24\micron\ excesses, and has colors similar to stars 
surrounded by optically thick, primordial disks (Paper I; \citealt{Dahm09}). 

The second population of 24\micron\ excess sources, present among all spectral
types, has weak or no detectable 8\micron\ excesses. Among the K0-M5 stars, the
magnitude of the 24\micron\ excess for this second population is less than the
24\micron\ excess found in any of the sources with both 8\micron\ and
24\micron\ excesses. If we assume the dust is isothermal, a lower limit to the
dust temperature obtained from the 24\micron\ excess and the 70\micron\ upper
limit is \about 53~K for blackbody emission. The upper limit to the fractional
dust luminosity for this dust temperature and a stellar temperature of 4000~K
is \about 10$^{-4}$. This fractional luminosity is orders of magnitude lower
than that found from optically thick disks around classical T Tauri stars. 

With the large number of \Spitzer\ surveys of star forming regions, the
primordial disk classification has been refined to include ``transitional''
\citep{Strom89}, ``pre-transitional'' \citep{Espaillat07}, ``evolved''
\citep{Hernandez07a}, ``homologously depleted'' \citep{Currie09a}, and
``anemic'' \citep{Lada06} designations. Each of these disk types is
characterized by reduced levels of emission at wavelengths \aboutless
10\micron\ relative to classical T Tauri stars. The reduced infrared emission
can be caused by a number of processes, including dynamical clearing of dust by
a companion, grain growth, and disk settling.

To compare the characteristics of the Upper Sco sample with anemic disks,
Figure~\ref{fig:ic348} shows the infrared colors from 
(\citealt{Lada06}, see also \citealt{Currie09b}) for
confirmed members of IC~348 \citep[see][]{Luhman03}. IC~348 was chosen for this
comparison since the cluster contains a similar range of spectral types as
Upper Sco, and was used to define anemic disks. One limitation, however, is
that the IC~348 observations detected the photosphere at 24\micron\ for A-type
stars only, and are not sensitive to the weak 24\micron\ excesses observed
around the later type stars in Upper Sco. The vertical dashed lines show the
range of slopes in the IRAC bands used to define anemic disks, while an
8\micron\ to 4.5\micron\ flux ratio greater than 0.64 would be classified as an
optically thick disk \citep{Lada06}. None of the B/A/F/G stars in Upper Sco
with an infrared excess would be classified with anemic or optically thick disk
based on these criteria. The circled crosses in Figure~\ref{fig:ic348} show the
actual stars in IC~348 that were identified by \citet{Lada06} to have anemic
disks based on the fitted slope to all 4 IRAC bands. For B7-A9 (top panel in
Figure~\ref{fig:ic348}) and F/G spectral types (middle panel), the anemic and
optically thick disks in IC~348 have larger 8\micron\ excesses than any of the
sources in Upper Sco. The 24\micron\ excesses also tend to be larger in IC~348,
although the distribution overlaps significantly with Upper Sco.

The Upper Sco sources with excesses detectable at 24\micron\ only also do not
fit the operating definition of transitional disks. Transitional disks are
noted by a lack of continuum excesses at wavelengths \aboutless 10\micron, but
retain excesses at longer wavelengths comparable in strength to classical T
Tauri stars. As shown in Figure~\ref{fig:ic348}, the K/M stars in Upper Sco
with excesses at 24\micron\ only have systematically lower 24\micron\
excesses than the stars with optically thick disks. This does not necessarily
indicate that the dust is debris, as a spectrum of inner hole sizes may exist,
and the weak 24\micron\ excesses may simply indicate a larger inner hole is 
present in the disks around the Upper Sco sample. While none of these systems 
were detected at 70\micron, most primordial disks were not detected at
70\micron\ either. Therefore, the 70\micron\ limits for the weaker 24\micron\
excesses are insufficient to rule out a cold, optically thick outer disk.

In summary, the above discussion suggests that Upper Sco contains two
populations of excess sources. The late type stars with strong 8\micron\ and
24\micron\ excesses plausibly have primordial disks based on the presence of
accretion signatures in a few stars, and a similarity in the infrared colors
of young stars surrounded by optically thick disks (Paper~I, \citealt{Dahm09}).
Many of these sources have infrared colors similar to ``anemic'' disks
\citep{Lada06}. The origin of the dust producing the weak 24\micron\ excesses
remains uncertain, but these disks are likely not a simple extension of the
transitional or anemic disk designations. Without additional observations, we
can only assume that these disks are gas-poor and the dust originates from a
debris disk, but we cannot rule out that the dust is remnant primordial
material.

In Table~\ref{tbl:excess}, we classify each of the disks candidates as
primordial, debris, or Be. Primordial disks were assigned to K/M type stars
that have 8\micron\ and 16\micron\ excesses since their infrared excess
characteristics are similar to optically thick disks \citep[see also][Paper
I]{Dahm09}. Two stars (HIP~77859 and HIP~78207) have prominent hydrogen
emissions in optical spectra and have been identified as classical Be stars
\citep{Crampton68,Jaschek64}; the observed infrared excess may originate from
free-free emission in an ionized gas disk \citep{Parter03}. The remaining
sources with infrared excesses were classified as debris disks given the
relatively weak excesses at 8, 16, or 24\micron. As noted in Paper~I, optically
thick primordial disks are present in 19\% of the K/M stars in the Upper Sco
sample, and in none of the B, A, F, and G stars. Debris disks are detected
around all spectral types, with percentages of 9\% for K/M stars, 13\% of F/G
stars, and 28\% of B/A stars. 

\section{Properties of the Debris Disks}
\label{prop}

A direct comparison of the debris fractions between early and late type stars
does not necessarily inform how the frequency of disks varies with spectral
type. Since observations at a given wavelength probe dust at smaller orbital 
radii around cooler stars for a fixed grain size, the low 24\micron\
debris fraction around K/M stars relative to A-stars may simply reflect that
the debris is located at large orbital radii and is too cool to detect in the
24\micron\ band. 

Nominally, the orbital radius of the debris can be inferred from the dust
temperature. This is possible only for the debris disks around the B/A stars
where the debris emission is detected at two or more wavelengths. For the
F/G/K/M stars, the debris emission is detected only at 24\micron, and a broad 
range of temperatures and radii can fit the single photometric point. We 
instead pose the question: can the debris properties inferred around the B/A 
stars also explain the debris emission observed around the later 
spectral types?

The emitted radiation from a debris disk varies with spectral type due to
differences in stellar heating and radiative ``blowout'' of small grains
\citep{Plavchan09}. To compare the debris properties between spectral types in
a self-consistent manner, we assume the debris is located in a narrow ring at
an orbital radius $R$. The particles follow a power-law size distribution,
$n(a)$, between grain radii $a_{\rm min}$ and $a_{\rm max}$ such that
\begin{equation}
  \label{eq:size}
  n(a) = N_{\rm o} \Bigl({a\over a_{\rm min}}\Bigr)^\alpha,
\end{equation}
where $N_{\rm o}$ is the normalization constant that effectively determines
the total grain surface area. The power-law exponent is fixed at $\alpha=-3.5$
as appropriate for an infinite collisional cascade without binary conditions on
the minimum and maximum particle size \citep{Dohnanyi69}. In practice, the
maximum grain size was fixed at $a_{max}=1000$\micron, and the minimum grain
size ($a_{\rm min}$) was set to either the blowout size or 0.05\micron, which
ever was larger. Excluding smaller grains does not impact the results
significantly since such small grains are inefficient emitters at 24\micron.

The grain sizes that will remain gravitationally bound to a star were
assessed by balancing the stellar gravitational force ($F_{\rm g}$) against 
repulsive forces from radiation ($F_{\rm pr}$) and stellar winds ($F_{\rm sw}$).
The ratio of these forces ($\beta$) for a particle of radius $a$ is given by
\begin{eqnarray} 
   \label{eq:beta}
   \beta(a) & = & {F_{\rm pr} + F_{\rm sw} \over F_{\rm g}}\nonumber \\
         & = & {3 L_* (Q_{\rm rad}(a) + Q_{\rm sw}(a)\,\dot{M}_{\rm sw}\,v_{\rm sw}\,c/\,L_*) \over 16 \pi G M_* \rho c}
\end{eqnarray} 
where $L_*$ is the stellar luminosity, $\dot{M}_{\rm sw}$ is the stellar mass
loss rate, $v_{\rm sw}$ is the stellar wind velocity, $Q_{\rm rad}(a)$ is the
particle cross section to radiation in units of the geometric cross section,
$Q_{\rm sw}(a)$ is the analogous cross section for stellar winds, and $\rho$ is
the grain density \citep{Burns79,Gustafson94,Strubbe06}. Assuming the disk
is optically thin, particles for which $\beta < 0.5$ remain gravitationally
bound to the star, while particles with $\beta > 0.5$ are ejected. The stellar
wind velocity was set to the escape velocity ($v_{\rm sw} = \sqrt{2GM_*/R_*}$).
The stellar mass loss rate is uncertain by orders of magnitude for these young
ages across all spectral types. The force from stellar winds is negligible for
the grain radii considered here unless the mass loss rate is \aboutmore 1000
times the solar value \citep{Plavchan05,Strubbe06}. We assume 
$\dot{M}_{\rm sw} = 2\times10^{-14}\,{M}_\odot$~yr$^{-1}$, which corresponds 
to the solar mass loss rate \citep{Wood04}, and that $Q_{\rm sw}(a)=1$. The
dust emission was computed using optical constants for astronomical silicates
\citep{Weingartner01} and assuming $\rho=2.7$~g~cm$^{-3}$. Stellar photospheres
were approximated as blackbodies.

Given the debris disk model, we can normalize the two free model parameters
($R$ and $N_{\rm o}$) using the nine B/A stars in Upper Sco that have both
16\micron\ and 24\micron\ excesses. The ratio of the 16\micron\ to 24\micron\ 
excess flux densities determines the orbital radius of the dust, $R$. The 
ratio varies between 2.8 and 10.0 with a median value of 3.8, which
leads to an orbital radius between 9~AU and 40~AU with a median of 15~AU. Given
the orbital radius, the total surface area of dust contained in particles
gravitationally bound to the star ($\beta < 0.5$) is set by the magnitude of
the 24\micron\ emission given the inferred orbital radius, which determines
$N_{\rm o}$ in Equation~\ref{eq:size}. The 24\micron\ excess for all 13 B/A
stars with debris disks varies between 0.5 and 16 times the photosphere, with a
median of \about 2. 

To compute the excess emission around stars of other spectral types, stellar
properties (mass, luminosity, and temperature) were obtained from the 5~Myr
pre-main-sequence isochrone of \citet{Siess00} for solar metallicity. The
orbital radius ($R$) and the size distribution ($N_{\rm o}$) of the dust were
fixed to that inferred around the B/A stars. However, for a given $L_*$ and
$M_*$, only grains with $\beta<0.5$ were included in the calculations of the
debris emission. Therefore, even though $N_{\rm o}$ is fixed, the total dust
surface area is larger for later spectral types since the radiation blowout
size is smaller. The ratio of the expected dust emission excess to the
photosphere was computed in this manner for stars between masses of 0.1\msun\
and 7\msun. 

Figure~\ref{fig:grains} presents the results of the calculations for 16\micron\
and 24\micron\ excesses. The shape of the model curves reflects variations in
stellar heating and the minimum grain size present in the disk. The dust
surface area decreases toward early spectral types and tends to reduce the dust
emission, which is compensated by warmer dust temperatures. The 24\micron\
excess relative to the photosphere (hereafter referred to as the relative
24\micron\ excess) peaks at a spectral type of \about B7 ($J-K_{\rm s} \approx
-0.06$), and declines toward more luminous stars as more small grains are
ejected by radiation pressure. Toward later spectral types, the relative
24\micron\ excess at first declines despite the increased surface area since
the dust temperature decreases with lower stellar heating. However, at a color
of $J-K_{\rm s} \approx 0.2$ (\about F0 star), the blowout particle radius is
\about 5\micron, and the dust temperatures are warmer than expected based on
blackbody radiation. Combined with the increased surface area of these smaller
grains, the normalized 24\micron\ excess ratio plateaus toward cooler stars
until a stellar color of $J-K_{\rm s} \approx 0.8$ (\about M0 star). At this
point the blowout size is less than 0.1\micron\ and these small grains emit
inefficiently at 24\micron. Therefore the additional surface area from small
grains does not appreciably increase the amount of infrared excess. The
16\micron\ curves exhibit similar behavior, but fall more steeply toward later
spectral types since the dust emission is further into the Wien regime and more
sensitive to the dust temperature.

Figure~\ref{fig:grains} shows that if the only variation in the debris
properties is the blowout size and stellar heating, the 24\micron\ excess
relative to the photosphere will be \about 2 times higher around \about A0
stars compared to F0-M0 stars at an age of 5~Myr. The relative 24\micron\
excess decreases rapidly toward stars earlier than B7 and later than M0.
Qualitatively, the range of relative 24\micron\ excesses around the
F/G/K/M-type stars are consistent with the extrapolation of the excesses around
the B/A stars based on this model. At 16\micron, the relative photometric
excess falls sharply with later spectral types and 16\micron\ excesses would
be undetectable toward the later type stars. This model is consistent with
the lack of 16\micron\ excesses from debris in F/G/K/M stars (Paper~I). We
conclude that is not necessary to invoke systematic differences
in the radial distribution of dust with spectral type to explain the range of 
observed 16\micron\ and 24\micron\ excesses.

\section{Evolution of Debris Emission}
\label{evolution}

Variations in the debris luminosity with stellar age provide a means to
investigate the evolution of planetesimal belts and potentially the formation
of planetary systems. \citet[see also \citealt{Wyatt07b}]{Dominik03} suggested
that the evolution of debris emission can be explained by quasi-steady state
collisional equilibrium of planetesimals distributed in a narrow ring. One 
limitation of this model is that it does not prescribe a physical mechanism to
initiate the collisional cascade. In a series of papers, \citet[and references
therein]{Kenyon08} presented a model that follows both the collisional growth
of planets and the onset of debris production. They show that the debris disk
initially has a low luminosity when planets are still in the formative stages.
When the collisional growth produces \about 2000~km size bodies, the resulting
gravitational interactions excite planetesimal collisions that leads to
dramatically increased debris luminosity. The timescale to form a 2000~km
body increases with the orbital radius, thereby producing a ``wave'' of debris
production in time that propagates from the inner disk outwards.

A prediction of the Kenyon \& Bromley model is that the debris luminosity
initially increases with time, and then declines once the wave of debris
production has propagated outward through the disk on a dynamical timescale.
For a debris disk that extends between 30 and 150~AU, the debris emission will
peak at an age of \about 5-30~Myr, with faster timescales for higher mass stars
for a fixed disk mass. A decline in debris luminosity for ages older than
\about 20~Myr has been observed in a number of studies and seems firmly
established by the data \citep{Habing01,Decin03,Rieke05,Siegler07,Carpenter09}.
\citet{Hernandez06} and \citet{Currie08a} further suggest that the 24\micron\
excess around A/F stars increases from an age of \about 5~Myr and peaks around
10-15~Myr. If verified, this initial increase in 24\micron\ debris luminosity
may signify the onset of planetesimal stirring in debris disks.

The Upper Sco observations presented here provide an important data point to
re-evaluate the trends at young ages since few surveys for debris disks are
available for younger stars than 10~Myr. As described in the Appendix, we have
compiled the results from a number of \Spitzer\ 24\micron\ surveys for
comparison to Upper Sco. Similar compilations have appeared in
\citet{Gaspar09}, \citet{Rebull08}, \citet{Currie08a}, and \citet{Siegler07}
among others. The compiled surveys include clusters and associations between
ages of 2 and 757~Myr, and main-sequence field stars between ages of 3 and
5000~Myr. The young associations and clusters that have been surveyed
previously are either sparsely populated with only \about 20 members
($\eta$~Cha, \citealt{Gautier08}; TW Hydra moving group, \citealt{Low05}) or
are at large distances such that \Spitzer\ is sensitive to the stellar
photosphere for only A-type stars or earlier (IC~348, \citealt{Lada06};
$\sigma$~Ori, \citealt{Hernandez07a}; Orion OB1a and OB1b,
\citealt{Hernandez06}; NGC~2262, \citealt{Currie08b}). 

The spectral type bins were selected to be B7-A9, F0-F9, and G0-K5, which
correspond to the spectral type ranges where numerous \Spitzer\ studies exist
(see Appendix). Stars earlier than B7 were omitted since they are fewer in
number, and as shown in Section~\ref{prop}, the debris luminosity may be
depressed by radiative blowout of the smallest grains. Stars later than K5 are
omitted since most \Spitzer\ surveys of young clusters and associations were
insensitive to the stellar photosphere for K/M stars. The relationship between
the empirical variable of spectral type and the physical variables 
stellar mass and luminosity, which dictate debris disk properties, will be 
discussed in Section~\ref{sens}.

As discussed in Section~\ref{nature}, we are faced with the ambiguity of
distinguishing if the ``anemic'', ``evolved'', and ``transitional'' disks 
are debris or primordial systems. We proceed by an entirely observational
definition. The disk around HR~4796A is commonly assumed to be debris in
nature. The ratio of the observed 24\micron\ emission to the photosphere for
this disk is 97, which is one of the largest relative 24\micron\ excesses 
known. We therefore assume that any B/A/F star with an ``anemic'', ''evolved'',
or ``transitional'' classification and has a observed-to-photospheric 
24\micron\ flux ratio of $\le 100$ is a debris disk, while more luminous 
systems are primordial disks. For G and K-type stars, we use a limit of 5 
based on the division shown in Figure~\ref{fig:ic348}. These divisions are 
arbitrary and not physically motivated. Changing these boundaries by even a 
factor of two has no substantial impact on our conclusions.

\subsection{24\micron\ excess fraction}
\label{evolution:fraction}

We first examine how the fraction of stars with 24\micron\ infrared excesses
varies with age. We consider two different thresholds to identify infrared
excess: 32\% to incorporate the largest number of studies, and 15\% to increase
sensitivity to fainter disks. The adopted thresholds for each survey are 
listed in the Appendix. Any surveys with sensitivity limits larger than
these thresholds were omitted from the appropriate plot. The excess fraction is
defined as the ratio of the number of debris disks to the total number of stars
(i.e. optically thick disks, debris disks, Be stars, and non-excess stars).
Stars were omitted from the statistics if the original study noted that the
photometry was contaminated (e.g. nebulosity, nearby bright star) such that is 
was not possible to determine if the star has an excess or not. 

Figures~\ref{fig:fraction24} presents the excess fraction for the three
spectral type groupings. For G0-K5 stars, the 24\micron\ excess fraction in 
Upper Sco is 11\% (6/54) at the 32\% detection threshold, which is 
indistinguishable from the excess fraction (6/56) observed for 10-30~Myr 
G0-K5 stars. This result is consistent with the FEPS program, which found that 
the excess fraction for solar-type stars is roughly constant for ages
\aboutless 300~Myr \citep{Meyer08,Carpenter09}.

Other studies have found that the debris disk fraction for solar-type stars can
be as high as 40-45\% between ages of 3 and 30~Myr  \citep{Siegler07,Gaspar09}.
The difference can be attributed to semantics, as these studies have defined
``solar-type'' stars as FGK-stars. Whether or not F-stars should be included in
the definition of solar-type stars is arbitrary. We treat them separately since
there is evidence that the excess fraction of F-type stars is higher than G0-K5
stars. In the 10-30~Myr bin, the excess fraction for F-type stars is 15/37
(41\%) at the 32\% detection threshold, compared to 6/56 (11\%) for G0-K5 star
in the same age bin. The probability that these two observed excess fractions
are drawn from the same parent population is $p=0.002$ according to the
two-tailed Fisher exact test. 

For B7-A9 and F0-F9 spectral types, the 24\micron\ excess fraction peaks at
\about 50\% for ages of \about 10-30~Myr and a 32\% detection threshold. The
possible peak in the excess fraction result was noted by \citet{Currie08b}, and
the decline in the excess fraction toward older ages was found by
\citet{Rieke05}. No such increase is apparent for the G/K stars.
\citet{Gaspar09} found that the excess fraction for F-type stars was greater
for ages than 3-10~Myr than for 10-30~Myr, which can be attributed to the high
excess fraction for F-type stars in Orion~OB1b and Orion~OB1a
\citep{Hernandez06}. Since \citet{Hernandez06} estimate that the Orion
24\micron\ observations are complete for stars earlier than F0 in Orion~OB1a
and A8 in Orion~OB1b, the excess fraction for F-stars in Orion may be biased
toward higher values and are not included in this analysis.

To evaluate if the excess fraction in the $<$10~Myr and 10-30~Myr age bins
could have been drawn from the same parent population for earlier spectral
types, we compared the ratios using the two-tailed Fisher's exact test. For a
24\micron\ excess threshold of 32\%, which has the largest sample of stars, the
probability that these two age bins are drawn from the parent population
is 17\% for both B7-A9 and F0-F9 spectral types. If we group B7-F9 spectral
types together, the corresponding probability is 9\%. The significance of the
enhancement in the excess fraction for 10-30~Myr stars relative to $<$10~Myr
stars then is \about 1.7$\sigma$ for stars earlier than F9. 
Therefore, we conclude that no strong evidence exists for a change in the
excess fraction from 2-10~Myr to 10-30~Myr among early-type stars.

\subsection{24\micron\ excess}
\label{evolution:lum}

Evolution in the debris disk properties may manifest itself in the excess
luminosity as well as the excess fraction. Figure~\ref{fig:excess24} shows the
ratio of the 24\micron\ excess to the photospheric flux as a function of
stellar age. Only stars brighter than the photospheric detection limit for the
respective samples are shown (see Appendix). Different symbols are shown for
primordial (asterisks) and debris (solid circles) disks; 3$\sigma$ upper 
limits are shown for stars without a detected excess (open triangles).

For B/A spectral types, the upper envelope of 24\micron\ excess emission
declines with time. \citet{Rieke05} suggested that for ages older than 5~Myr,
the upper envelope varies with stellar age as $t^{-1}$, which is shown as the
dashed line in the top panel of Figure~\ref{fig:excess24} and extrapolated to
younger ages. The relative 24\micron\ excesses observed around B/A stars in
Upper Sco are consistent qualitatively with this upper envelope. For G0-K5
spectral types, the peak relative 24\micron\ excess is roughly constant for
ages less than 300~Myr, with a possible decline toward older ages. For F-type
stars, a sharp peak in the 24\micron\ excess in present at an age of \about
16~Myr as previously noted by \citet{Hernandez06} and \citet{Currie08a}. This
peak is due to luminous debris disks in the Lower Centaurus Crux association
\citep{Chen05}.

To evaluate the trends quantitatively, we used the generalized Kendall's Tau
test as implemented in the ASURV Rev 1.2 package \citep{Lavalley92}. This test
incorporates both the detections and upper limits to make efficient use of
the data. We first consider if evolution is present over all ages in the
sample. The probability that a correlation of the 24\micron\ excess versus age
is not present for B7-A9, F0-F9, and G0-K5 stars is $3\times10^{-10}$,
$2\times10^{-8}$, and $10^{-5}$, respectively. These results confirm the trend
apparent from visual inspection of Figures~\ref{fig:excess24} that the
magnitude of the 24\micron\ excess decreases with stellar age over all spectral
types considered here. This conclusion has been reached previously based on
\ISO\ \citep{Habing01,Spangler01}, \IRAS\ \citep{Moor06}, and \Spitzer\
\citep{Rieke05,Su06,Siegler07,Meyer08,Carpenter09} observations.

We now consider if evolution in the 24\micron\ excess is present at young ages.
We specifically consider the age range between 5 and 17~Myr, which is well
populated with clusters and associations, and significant evolution is
anticipated based on theoretical models for the collisional evolution of a
Kuiper Belt analog \citep{Kenyon08}. For B7-A9, F0-F9, and G0-K5 spectral
types, the probability that a correlation of the 24\micron\ excess versus age
is not present between 5 and 17~Myr is 0.14, 0.01, and 0.56, respectively. Thus
when subdivided by spectral type, the only suggestion of a trend in the
24\micron\ excess with age between 5 and 17~Myr is for F-type stars. From
inspection of Figure~\ref{fig:excess24}, the trend is such that the amount of
24\micron\ excess increases over this age range.

\citet{Currie08a} reported a significant rise in the magnitude of the debris
emission between 5 and 17~Myr based on debris disk observations around B/A/F
stars in Orion~OB1b, Orion~OB1a, LCC, and UCL. They used the Wilcoxon rank-sum
test to determine the probability ($p$) that the mean excess is consistent
between two samples. The strongest statistical trend identified by
\citet{Currie08a} was the increase in the mean 24\micron\ excess around A/F
stars in the 5~Myr Orion~OB1b association compared to the 8.5~Myr Orion~OB1a 
association ($p=0.002$), and Orion~OB1b compared to a combined 16-17~Myr old
LCC/UCL sample ($p=0.05$). One difference in the analysis conducted by
\citet{Currie08a} and this study is that we separated the F-type stars from the
B/A stars, and we did not include the F-type stars in Orion~OB1a and
Orion~OB1b. While the separation by spectral type is arbitrary from a physical
point of view, the observations are complete for B/A spectral types over a
broader range of ages. The MIPS~24\micron\ observations of Orion~OB1a and
Orion~OB1b in particular are complete to the photosphere for F0 and A8 stars
\citep{Hernandez06}, respectively, and the 24\micron\ detections of F-type
stars in these associations may be biased toward stars with disks. 

We stress that we have not included in this analysis the 17 debris disks
identified in the MIPS 24\micron\ survey of the 13~Myr old, double cluster h \&
$\chi$ Persei \citep{Currie08a}. These 17 stars have A/F spectral types with
relative 24\micron\ excesses that range between 8 and 180, with a median
value of 19. These excesses are extreme compared to the sources
shown in Figures~\ref{fig:excess24}. These luminous debris disks could
conceivably indicate a peak in the debris production in the 10-20~Myr age
range. However, the parent sample for the h \& $\chi$ Persei is about 30 times
larger than the Upper Sco sample. As noted by \citet{Currie08a}, the extreme
excesses in h \& $\chi$ Persei may simply reflect better sampling of the same
parent luminosity function. Deeper observations and tabulation of the measured
flux densities for all cluster members will determine if the extreme excess
sources in h \& $\chi$ Persei represent a peak in the debris production.

\subsection{Interpretation}
\label{sens}

Before drawing further conclusions from the results presented in
Section~\ref{evolution:lum}, we consider how the stellar properties may
influence the interpretation of debris disk evolution. The data presented in
Figures~\ref{fig:fraction24} and \ref{fig:excess24} were grouped by the
observed spectral type. For main sequence stars, the spectral type bins of
B7-A9, F0-F9, and G0-K5 correspond approximately to stellar masses of 1.9, 1.4,
and 0.8\msun. The corresponding Kelvin-Helmholtz contraction times are \about
3, 10, and 30~Myr. Since the ages of the stellar samples are as young as 2~Myr,
the younger stars are in the pre-main-sequence phase of evolution and the
stellar mass and luminosity will vary with age for each of the spectral type
bins. This is particularly true for young ($<$ 10~Myr) stars more massive than
\about 1\msun, which are evolving toward the main-sequence along the Henyey
tracks. For example, the spectral type for a 2\msun\ star will be K2, G8, and
A2 at an age of 2, 5, and 10~Myr, respectively according to the \citet{Siess00}
pre-main-sequence evolutionary models and \citet{Kenyon95} temperature scale.

Since the blowout size is proportional to $L_*/M_*$ (see
Equation~\ref{eq:beta}), the minimum grain size in the debris disk will also
vary with age. For a fixed spectral type the $L_*/M_*$ ratio, and hence the
blowout particle size, decreases by a factor of \about 5 between 1~Myr and
10~Myr for A0-M0 stars. The total grain surface area is proportional to $a_{\rm
min}^{-0.5}$ for a $a^{-3.5}$ particle size distribution, and the debris
luminosity will thus increase in time if the planetesimal belt is in otherwise
steady state.

The expected trends were evaluated more quantitatively using the debris disk
model described in Section~\ref{prop}. For a given age and spectral type (i.e.
stellar temperature), the stellar mass and luminosity were determined from the
\citet{Siess00} evolutionary models for solar metallicity and no convective
overshoot. The 24\micron\ debris emission was then computed for a planetesimal
belt at an orbital radius of 15~AU, with a minimum grain size corresponding to
the blowout size. The results are insensitive to the assumed orbital radius
since the variations in the debris luminosity with time are caused primarily by
changes in the grain surface area. We assume that the minimum grain size in
the debris disk varies instantaneously as stars evolve since the collisional
lifetime of dust within an orbital radius of 30~AU is less than 1~Myr for
observed debris disks \citep{Backman93}. In this manner, the variation in the
24\micron\ excess relative to the photosphere was computed considering only
stellar evolution. 

Figure~\ref{fig:sens} shows the results of the model calculations. For A0
through K0 stars, the relative 24\micron\ excess increases by a factor of
\about 5 from an age of 1~Myr to a peak at 10-50~Myr. The peak relative excess
is reached sooner around earlier spectral types since the stars reach the
main-sequence on shorter time scales. M-type stars obtain a peak relative
excess in only a few million years when the grain blowout size is less than
0.05\micron. At this point, any grains removed by radiation pressure are
inefficient emitters at 24\micron, and the debris luminosity declines with the
stellar luminosity. \footnote{ MIPS observations have yielded a small number of
debris disks around young, low mass T Tauri stars \citep{Cieza07}. The low
excess needs to be interpreted in context of these model calculations, which
predict the 24\micron\ excess will be 5 times weaker than a 10~Myr star if the
planetesimal belt is otherwise the same.} If the model assumptions are correct,
the sensitivity functions presented in Figure~\ref{fig:sens} will be
superimposed on the evolution of the debris luminosity and must be considered
when attributing any increase in the debris luminosity with age to planet
formation. 

For comparison, the gray solid line in Figure~\ref{fig:sens} shows the
variation in the 24\micron\ excess relative to the photosphere for a 2\msun\
star (\about A3 spectral type) with a planetesimal belt between 30 and 150~AU
from the \citet{Kenyon08} models. Since the stellar luminosity is held constant
in these models, the temporal variation in the infrared luminosity results from
changes in the debris production. The debris evolution in these models depends
on the stellar mass and the planetesimal belt properties (mass, orbital
radius), but the general prediction is that the fractional debris luminosity
rises at early ages, and declines at later times. The results show that for a
30-150~AU planetesimal belt, the rise in the debris luminosity and the age
where debris production is maximized resembles the evolutionary curve expected
based on stellar evolution alone. Clear differences are observed at older ages
where the debris luminosity in the \citet{Kenyon08} models declines more
rapidly.

We now evaluate if the magnitude of the relative 24\micron\ excess varies
with stellar age for ages $<20$~Myr, and if any increase can be
attributed to stellar evolution. For
both the B7-A9 and G0-K5 spectral type ranges, any variations in the
24\micron\ excess with age between ages of 5 and 17~Myr have a significance of
less than 2$\sigma$ (see Section~\ref{evolution:lum} and
Figure~\ref{fig:excess24}). Thus independent of the selection function
presented in Figure~\ref{fig:sens}, no compelling evidence exists for a
variation in the mean relative 24\micron\ excess for B7-A9 and G0-K5 stars
in that age range.

The F-type stars show the strongest evidence (2.6$\sigma$) for a rise in the
relative 24\micron\ excess (see Section~\ref{evolution:lum}), but these
results need to be interpreted in view of the expected increase in the debris
emission shown in Figure~\ref{fig:sens}. For the F-type stars between ages of 5
and 17~Myr, we scaled the magnitude of the 24\micron\ excess by the inverse of
the model calculations computed for a F5 star, which corrects for the reduced
excess emission expected to occur with stellar evolution. The probability of a
trend in the observed excess emission with age (see Figure~\ref{fig:excess24})
was then recomputed using the Kendall Tau test. We find the probability that a
correlation is not present increases from $p=0.01$ (2.6$\sigma$) to $p=0.02$
(2.3$\sigma$). 

In summary, we conclude that any evidence for an increase in the 24\micron\
debris emission with age between 5 and 17~Myr rests primarily with F-type
stars, and the significance is \aboutless 2.6$\sigma$. Any variations in the
relative 24\micron\ excess over this age range for B/A and G/K are significant
at less than 2$\sigma$ confidence. Given the low statistical significance and
that the increasing relative excess is observed in a narrow spectral type
range, we hesitate to ascribe the apparent rise in the mean relative
24\micron\ excess among F-stars to the onset of planetesimal stirring. 

\subsection{Debris disks around M-stars}

Evolution of debris emission around M-stars is more difficult to quantify since
most MIPS surveys are insensitive to the stellar photosphere for these spectral
types. Nonetheless, for completeness, we present in Figure~\ref{fig:mstars} the
24\micron\ excess relative to photosphere as a function of age for M0-M5 stars
in IC~348 \citep{Lada06}, $\sigma$~Ori \citep{Hernandez07a}, Upper Sco,
$\eta$~Cha \citep{Gautier08}, TW~Hydra moving group \citep{Low05}, NGC~2547
\citep{Forbrich08}, and IC~2391 \citep{Siegler07}. M-stars with upper limits to
the 24\micron\ excess are not shown for clarity, and because most studies have
not reported upper limits if a star is not detected. 

Figure~\ref{fig:mstars} again illustrates the dichotomy at young ages between
classified primordial disks (asterisks) and debris disks (filled circles). The
24\micron\ excess around Upper Sco M-stars is comparable to many of the stars
in NGC~2547 at an age of \about 38~Myr \citep{Naylor06}. However, 5 M-stars in
NGC~2547 have a 24\micron\ excess of \about 10 times the photosphere; three of
these 5 stars have been observed with IRAC \citep{Young04}, and one has an
8\micron\ excess. By contrast, all of the Upper Sco sources classified with a
primordial disk have an 8\micron\ excess. It would be unusual for an optically
thick disk to persist for \about 38~Myr, and the debris disk is a reasonable
interpretation for the NGC~2547 sources. These results suggest that
distinguishing primordial disks from debris disks for M-type stars merits a
more detailed examination \citep[cf.][]{Ercolano09}.

\section{Summary}

We have presented the results of a \Spitzer\ 24\micron\ and 70\micron\ 
photometric survey of 205 members of the Upper Scorpius OB association.
These data were combined with MIPS photometry for 15 Upper Sco sources 
observed by the FEPS Legacy program to provide a census of circumstellar disks 
around stars with spectral ranges ranking from B0 to M5 at an association age
of 5~Myr \citep{Preibisch02}. By analyzing the 24\micron\ photometry with
4.5\micron\ and 8\micron\ measurements presented in Paper~I, we
identify 54 stars that have observed 24\micron\ emission that exceeds the
expected stellar photospheric emission by 32\% or more. Similarly, 12 stars
were identified with a $\ge3\sigma$ photometric excess at 70\micron; all 12 of
these stars have a detectable 24\micron\ excess. 

The nature of the excess sources were established based on the color and
magnitude of the excess emission. We find a dichotomy of excess
characteristics. One population, found only around the K and M-type stars, has
strong excess emission at both 8\micron\ and 24\micron\ that is comparable to
known optically thick circumstellar disks. A second population, found around
all spectral types, has weak 8\micron\ and 24\micron\ excesses compared to
optically thick and ``anemic'' disks in the IC~348 cluster \citep{Lada06}. We
suggest that these weak excesses originate from debris disks formed from the
collisional shattering of planetesimals, although we cannot exclude the
possibility that these systems are the remnants of optically thick disks. Of
the 54 excess stars, we attribute the 24\micron\ excess emission to 2 Be stars,
24 primordial disks, and 28 debris disks. The debris disks include 11 K/M
stars.

The debris disks were analyzed to investigate whether the orbital radius of the
presumed planetesimal belts vary systematically with spectral type. We
modeled the emission with a power-law distribution of grain sizes following
the \citet{Dohnanyi69} equilibrium distribution ($n(a) \propto a^{-3.5}$).
The orbital radius ($R$) and grain size distribution ($N_{\rm o}$; see 
Equation~\ref{eq:size}) were set to reproduce the observed excess emission
around nine B/A stars with infrared excesses at both 16\micron\ and 24\micron.
The expected emission around stars of other spectral types were then estimated
after allowing for variations in the stellar heating and radiation blowout of
small grains. We find that the magnitude of the 24\micron\ emission observed
around later stars can be produced by this model, indicating it is not
necessary to invoke a different radial distribution. This model predicts a
steep fall off in the relative 24\micron\ excess for stars earlier than
\about B7 due to blowout of the small grains, and later than \about M0 due to
reduced stellar heating.

The Upper Sco results are combined with other \Spitzer\ 24\micron\ surveys in
the literature to reassess the evolution of 24\micron\ debris emission. After
consideration of both sample sizes and detection limits, we find a decline in
the magnitude of the 24\micron\ excess relative to the photosphere for spectral
types between B7 and K5 and ages between 10~Myr and 5~Gyr, as has been noted in
previous studies \citep{Habing01,Decin03,Rieke05,Carpenter09}. We also
investigated if the 24\micron\ excess increases with stellar age for ages
\aboutless 20~Myr. Such an increase may be indicative of the onset of 
planetesimal stirring \citep{Kenyon08}. We caution, however, that a similar 
increase in the excess emission may also result from pre-main sequence
evolution; for a fixed spectral type, the $L_*/M_*$ ratio will decrease with
age, which leads to a higher debris luminosity as fewer grains are ejected by
radiative forces. The observed mean 24\micron\ excess around F-type stars
increases between ages of 5 and 17~Myr as previously found by \citet[see also
\citealt{Hernandez06}]{Currie08a}, but with a significance of \about
2.3-2.6$\sigma$ confidence. The variations in the mean 24\micron\ excess around
B/A and G/K stars over the same age range are significant at less than
2$\sigma$ confidence. We conclude that the observational data do not yet
require a physical mechanism to produce a peak in the observed debris emission.

\acknowledgements

We are grateful to Scott Kenyon for his comments on the manuscript. This work
is based on observations made with the {\it Spitzer Space Telescope}, which is
operated by JPL/Caltech under a contract with NASA. Support for this work was
provided by NASA through an award issued by JPL/Caltech. This research made use
of the SIMBAD database, the IRSA archive, and the Two Micron All Sky Survey.

\appendix

\section{Compilation of \Spitzer\ Debris Disk Surveys}

To compare the Upper Sco results with previous observations, we compiled data
from published \Spitzer\ 24\micron\ surveys for debris disks. We considered
studies in which the parent sample is not biased for or against the presence of
the debris disk. This requirement implies that the parent sample must have been
selected by photospheric indicators, and that the \Spitzer\ observations were
sensitive to the stellar photosphere. In practice this is not strictly true for
nearby moving group such as $\beta$~Pic, $\eta$~Cha, and TW~Hydra, where the
moving group was initially identified based on the presence of an apparently
isolated sources with an prominent infrared excess. We nonetheless include
these groups since the subsequent surveys for moving group members were not
biased toward circumstellar disks.

Table~\ref{tbl:lit} summarizes the surveys that were compiled and contain 5
stars or more. We include in this table the name of the region, the adopted
age, the earliest spectral type in the survey, the latest spectral type at
which the photosphere can be reliable detected as discussed above, the number
of stars in this spectral type range, and the minimum 24\micron\ excess above
the stellar photosphere that could be detected. In some regions (e.g. NGC~2232,
Pleiades, Praesepe), not all stars have spectral types available from
spectroscopic observations, and the spectral type was estimated from optical or
near-infrared colors. Field stars with ages between 3~Myr and 5~Gyr were 
compiled from the FEPS survey of solar-type stars \citep{Meyer06}, the survey
of solar-type field stars by \citet{Beichman06} and \citet{Trilling08}, and the
survey of B/A stars by \citet{Rieke05} and \citet{Su06}.

The photometry was extracted from the publications listed in
Table~\ref{tbl:lit}. No attempt was made to re-reduce the data under a common
data reduction method or to re-assess cluster or association membership.
Several studies noted sources where the photometry was contaminated by
nebulosity or a companion sources. If these stars were excluded from the
analysis in the original study, they were also excluded in our analysis. After
compiling the source lists, we ran all source names through SIMBAD to identify
duplicate sources appearing under different names. If photometry appear for a
star in more than one survey, we adopted the photometry from the survey with
the higher signal to noise.

\begin{figure}
\begin{center}
\includegraphics[angle=-90,scale=0.8]{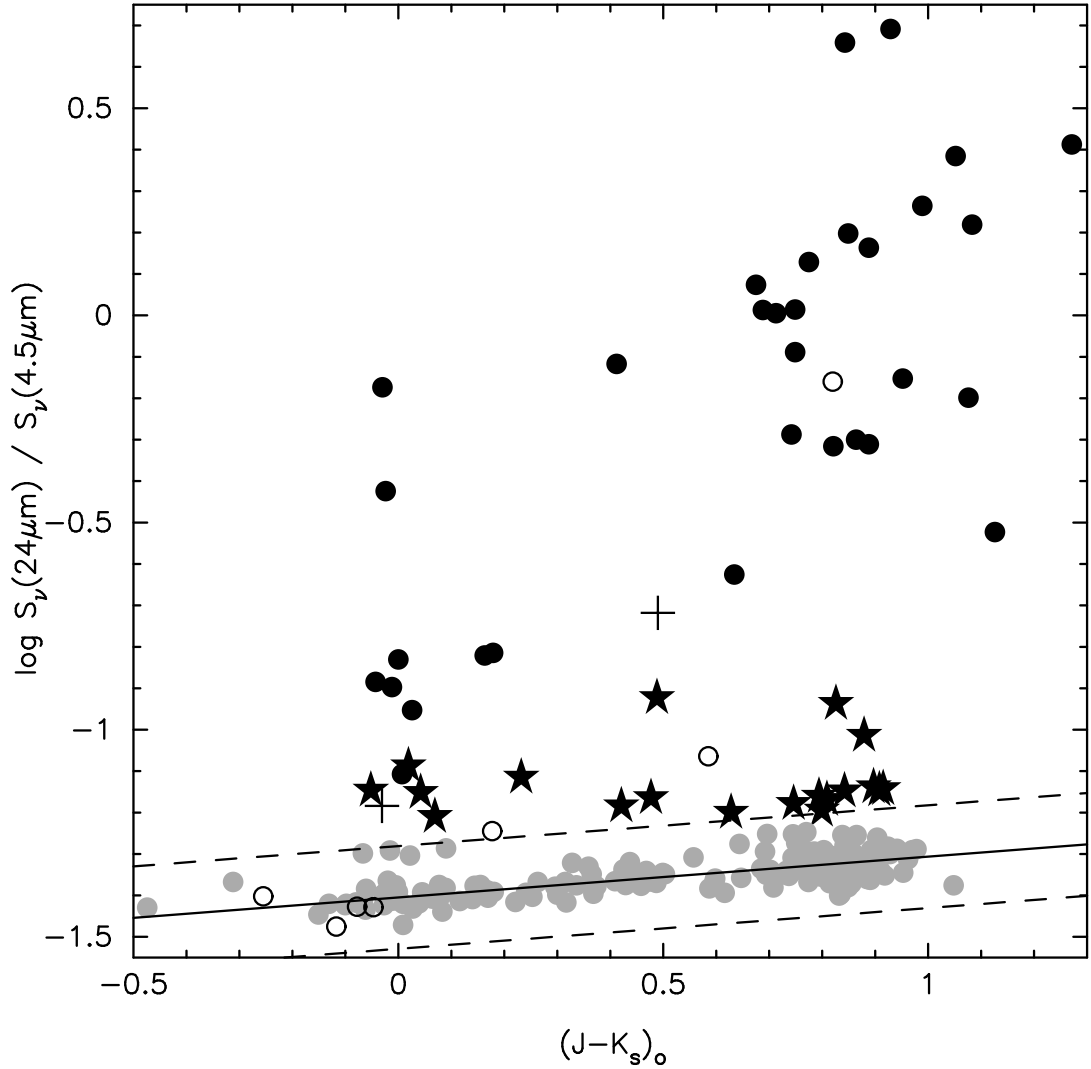}
\caption{
  \label{fig:ccd} 
  24\micron\ to 4.5\micron\ flux density ratio (\RM) as a function of
  dereddened \JK\ color. The solid line indicates the best-fit line [log \RM\ = 
  0.098 \JKO $- 1.40$] to the main locus of points that is assumed to 
  represent the stellar photosphere. The dashed lines indicate the 
  $\pm4\sigma$ limits about the fit ($\sigma=0.03$~dex). Black circles 
  indicate sources with an 8\micron\ or 16\micron\
  excess from Paper~I. Black stars are sources with detectable excesses at
  wavelengths $\ge 24$\micron, while gray circles indicate sources where the
  colors are consistent with photospheric emission. Sources with questionable
  photometry from nebulosity or a contaminating source are indicated by open
  circles, and known Be stars are marked by the crosses. The source
  [PBB2002] USco J161420.2$-$190648 is offscale on this plot at 
  \JKO,\RM = 2.14, 0.27. 
}
\end{center}
\end{figure}

\begin{figure}
\begin{center}
\includegraphics[scale=0.8]{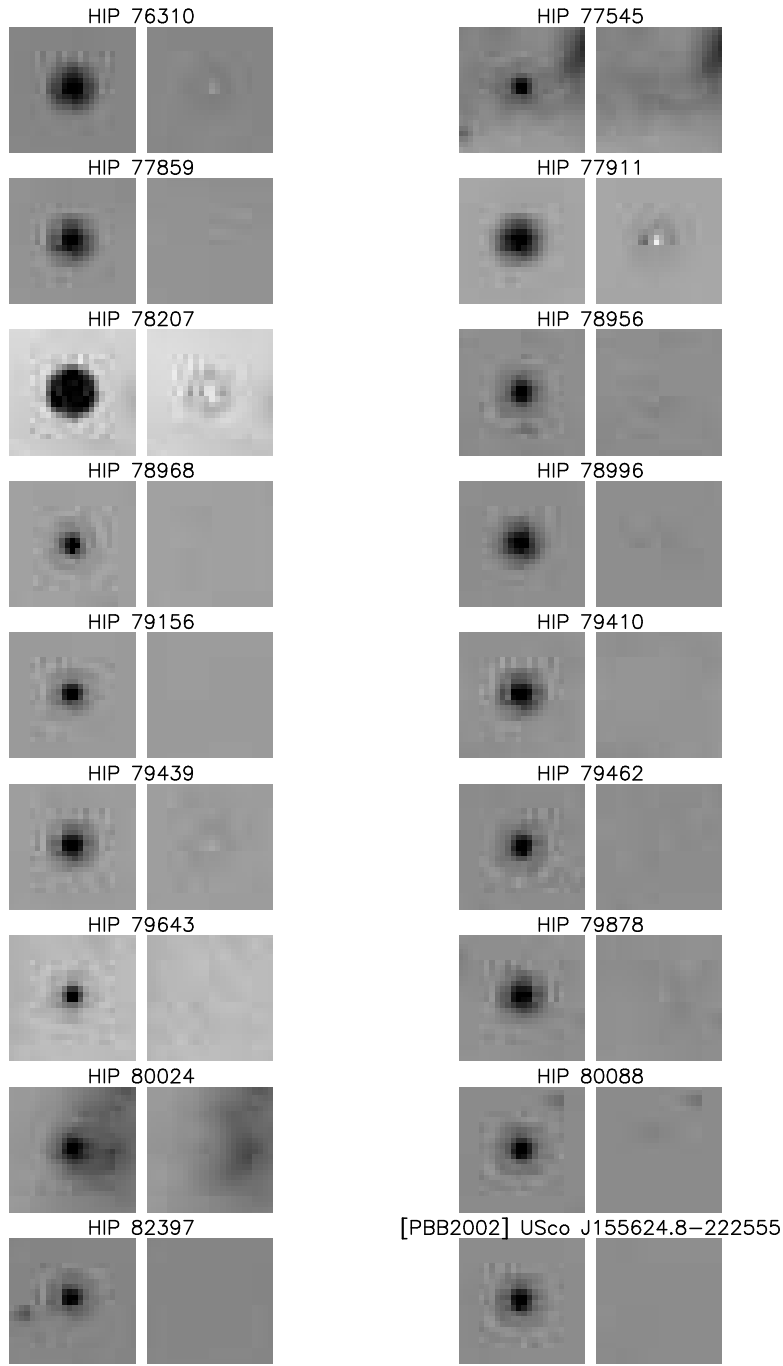}
\caption{
  \label{fig:images} 
  24\micron\ images for the 54 Upper Sco stars with a 24\micron\ excesses. 
  The left panel for each star 
  shows a $1'\times1'$ region centered on the stellar position.
  The right panel shows the PRF-subtracted image, where all identified stars 
  in the field of view have been fitted and removed. All images are displayed 
  in a logarithmic scale.
}
\end{center}
\end{figure}

\begin{figure}
\begin{center}
\includegraphics[angle=-90,scale=0.9]{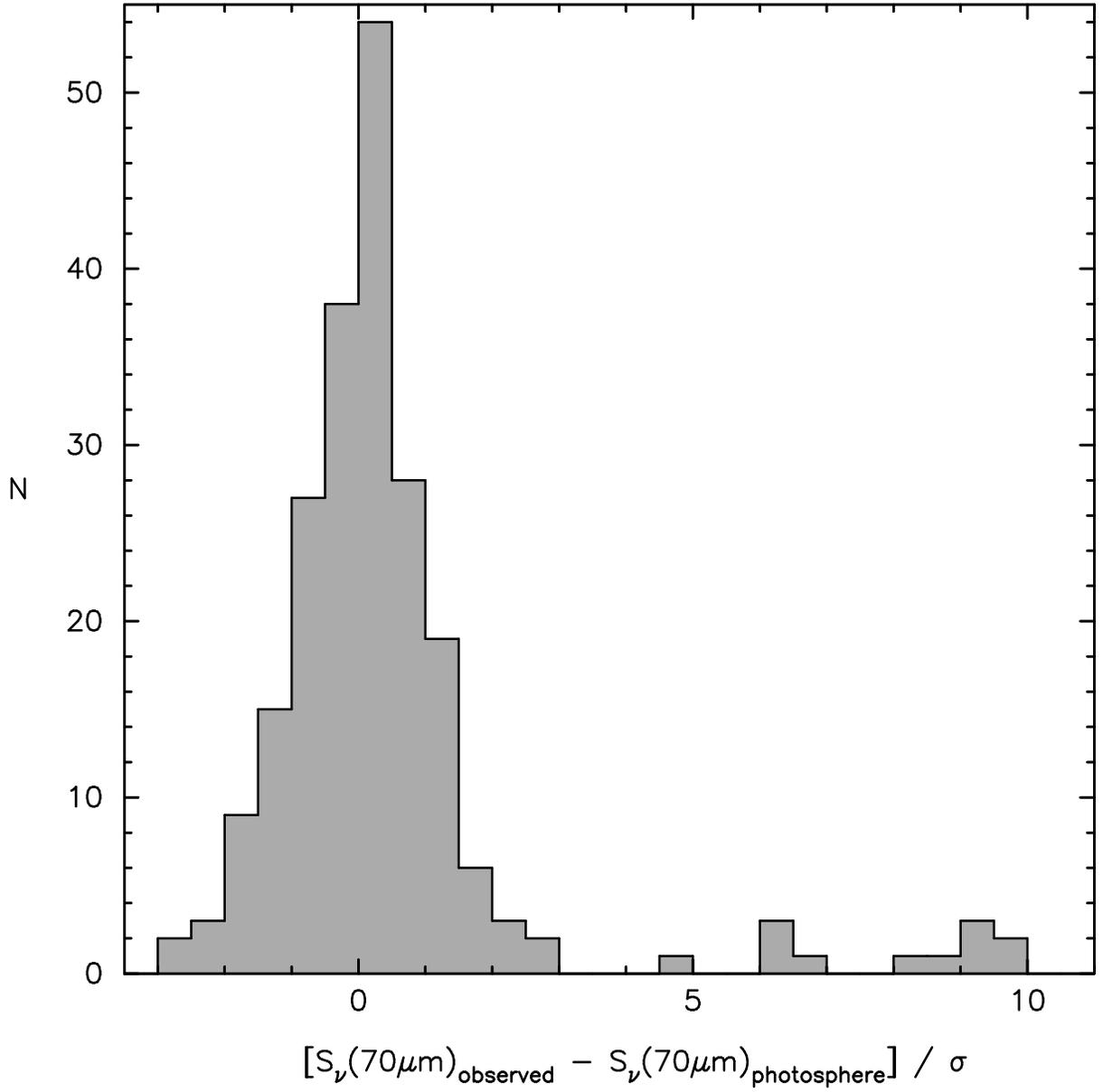}
\caption{
  \label{fig:mips70_excess}
  Histogram of the signal-to-noise ratio for the observed 70\micron\ excess 
  above the stellar photosphere. The histogram excludes two stars (ScoPMS 17
  and HIP~80338) for which the 70\micron\ photometry is suspect (see 
  Section~\ref{obs:mips70}). Sources were identified as having a significant 
  excess if the observed 70\micron\ flux density exceeded the
  photospheric flux density by $\ge3\sigma$.
}
\end{center}
\end{figure}

\begin{figure}
\begin{center}
\includegraphics[scale=0.8]{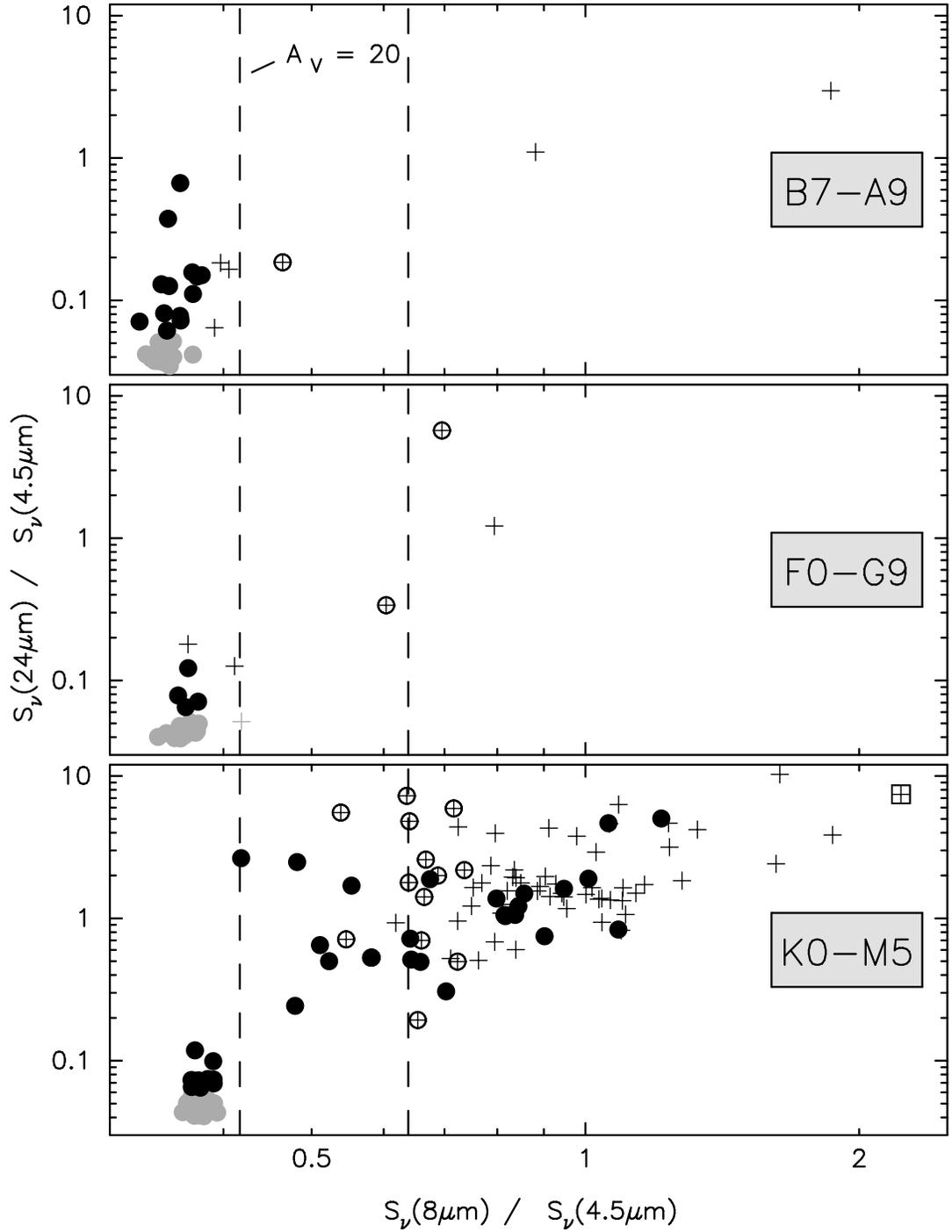}
\caption{
  \label{fig:ic348} 
  The dereddened 24\micron\ to 4.5\micron\ flux ratio as a function of the
  dereddened 8\micron\ to 4.5\micron\ flux ratio for B7-A9 stars (top),
  F0-G9 stars (middle), and K0-M5 stars (bottom). Filled circles represent 
  stars
  in Upper Sco, and crosses indicate stars in IC~348 that have a 24\micron\
  detection \citep{Lada06}. Black symbols indicate stars with an infrared
  excess at 24\micron\ (excluding two Be stars), and gray symbols indicate
  stars that lack a 24\micron\ excess. Crosses bounded by circles and squares 
  are stars in IC~348 that have been classified by \citet{Lada06} as ``anemic'' 
  or ``flat'' based on the slope of the SED in the 
  \Spitzer\ IRAC bands. The dashed lines indicate the range of slopes 
  ($\alpha = -2.56$ to 1.80) used to define anemic disks 
  ($\alpha = d {\rm log}(\nu S_\nu) / d {\rm log}(\nu)$). A few anemic disks 
  lie outside these boundaries since the actual slopes were computed using all 
  4 IRAC bands. A G0 star in IC~348 is offscale at (6.4, 4.8). 
}
\end{center}
\end{figure}

\begin{figure}
\begin{center}
\includegraphics[angle=0,scale=0.9]{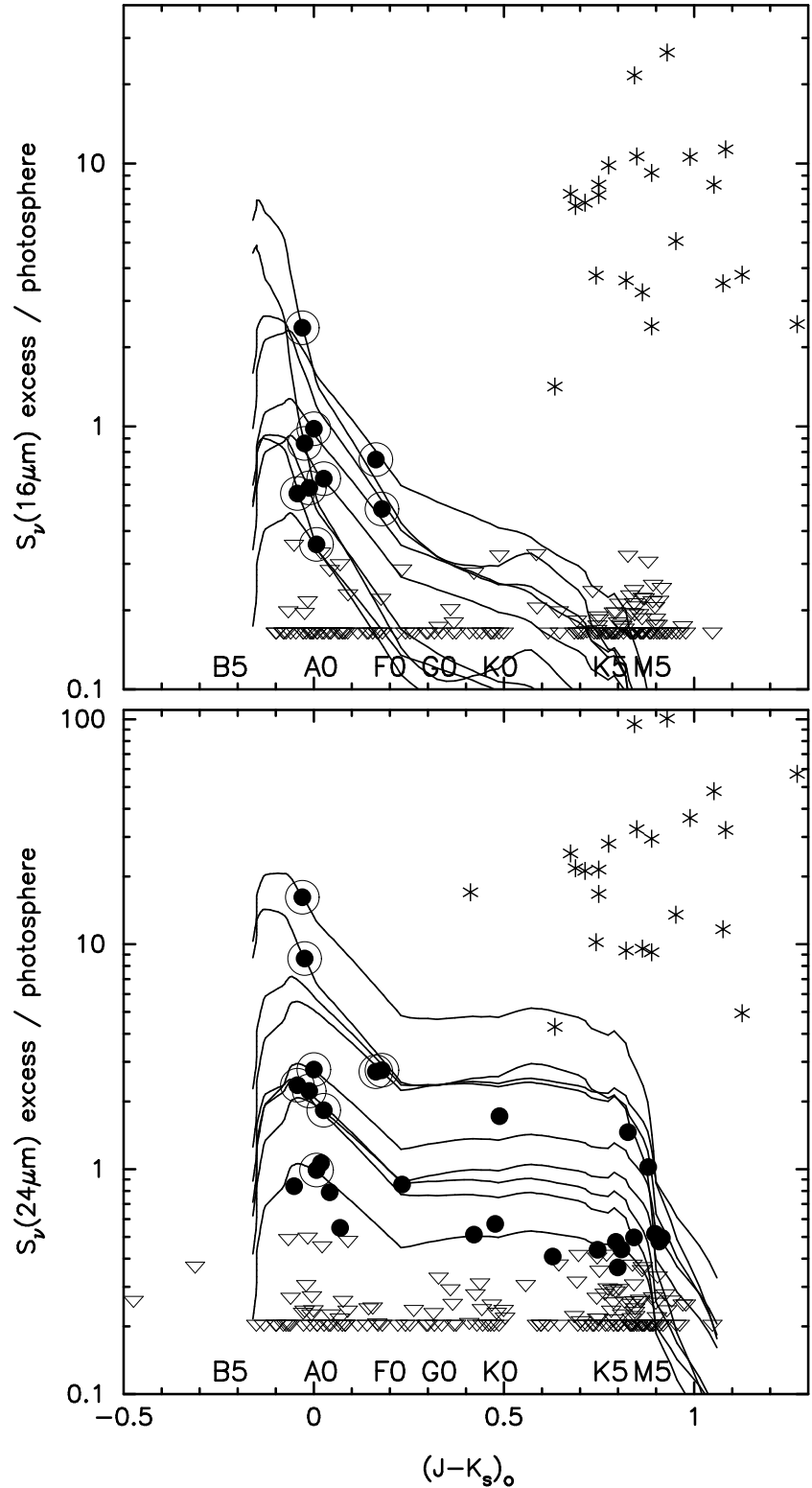}
\caption{
  \label{fig:grains} 
  The 16\micron\ (top panel) and 24\micron\ (bottom panel) excess in Upper Sco 
  normalized by the photospheric emission
  as a function of the dereddened $J-K_{\rm s}$ color. Symbols indicate debris 
  (filled circles) and primordial (asterisks) disks. Triangles indicate 
  the 3$\sigma$ upper limits for stars without detected excesses. Be stars
  and sources with contaminated photometry have been omitted.
  The solid curves show the emission from model debris disks with a power-law 
  distribution of particle sizes at a given orbital radius. The models
  were normalized to the debris properties around
  B/A stars, and computed around stars of various temperatures after 
  factoring differences in stellar heating and radiation blowout size.
  Open circles indicate the nine B/A stars used to normalize the models.
  See text for explanation of model behavior as a function of photospheric
  color. The results indicate that the magnitude of the 16\micron\ and
  24\micron\ excess around the debris disks can be explained by planetesimals 
  belts that have a similar range of orbital radii across all spectral types.
}
\end{center}
\end{figure}

\begin{figure}
\begin{center}
\includegraphics[scale=0.7]{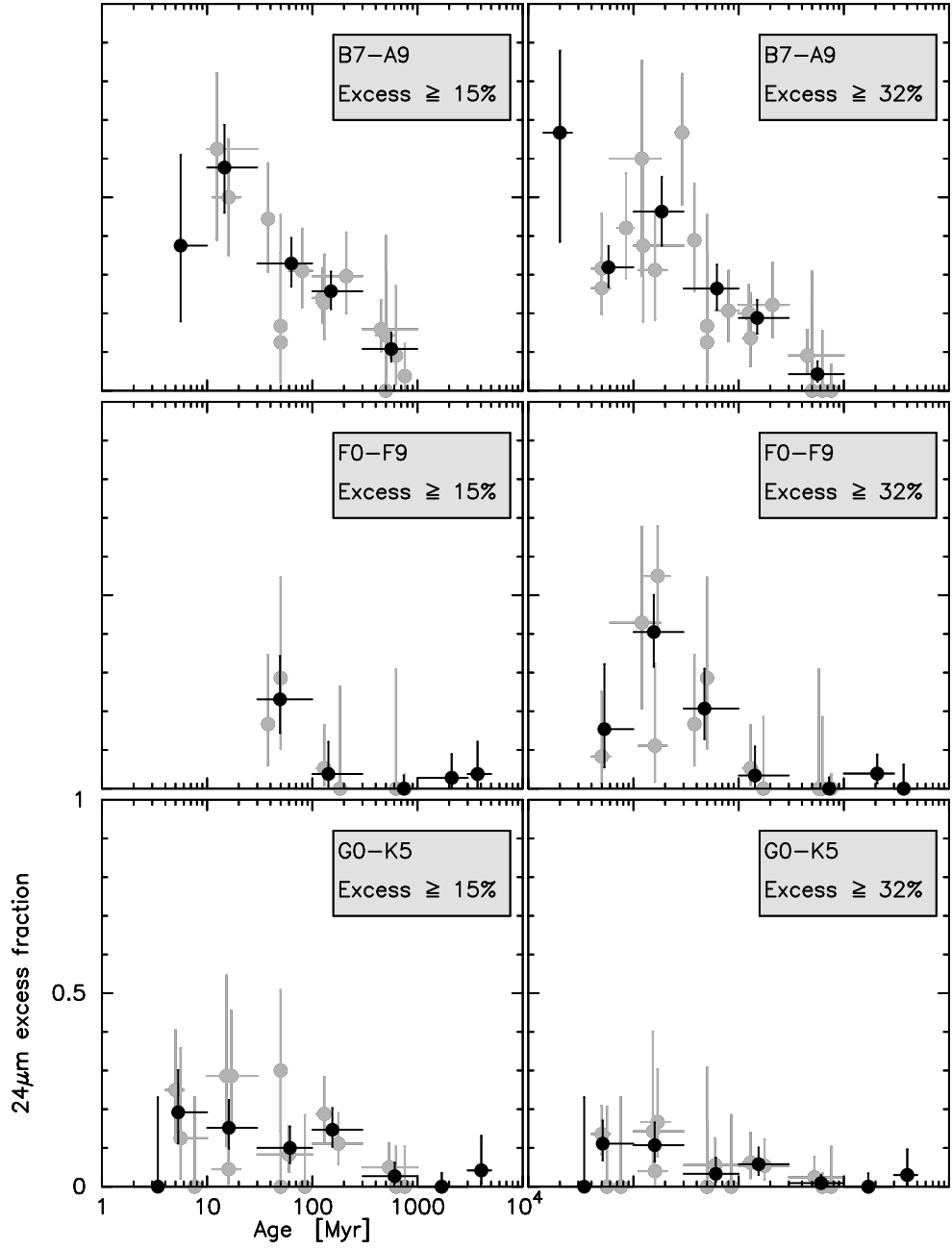}
\caption{
  \label{fig:fraction24}
  Fraction of sources with a 24\micron\ excess from a debris disk versus 
  stellar age for the regions summarized in Table~\ref{tbl:lit}. The plots 
  in the left 
  panels show the results for a 24\micron\ excess detection threshold of 
  15\%, and the right panels for 32\%. Results are presented for three
  spectral type ranges. Grey circles represent individual clusters or 
  associations that contain at least 5 stars in the appropriate mass range. 
  Black circles represent the ensemble average of field stars, clusters, 
  and associations regardless of the sample size.
}
\end{center}
\end{figure}

\begin{figure}
\begin{center}
\includegraphics[scale=0.7,angle=0]{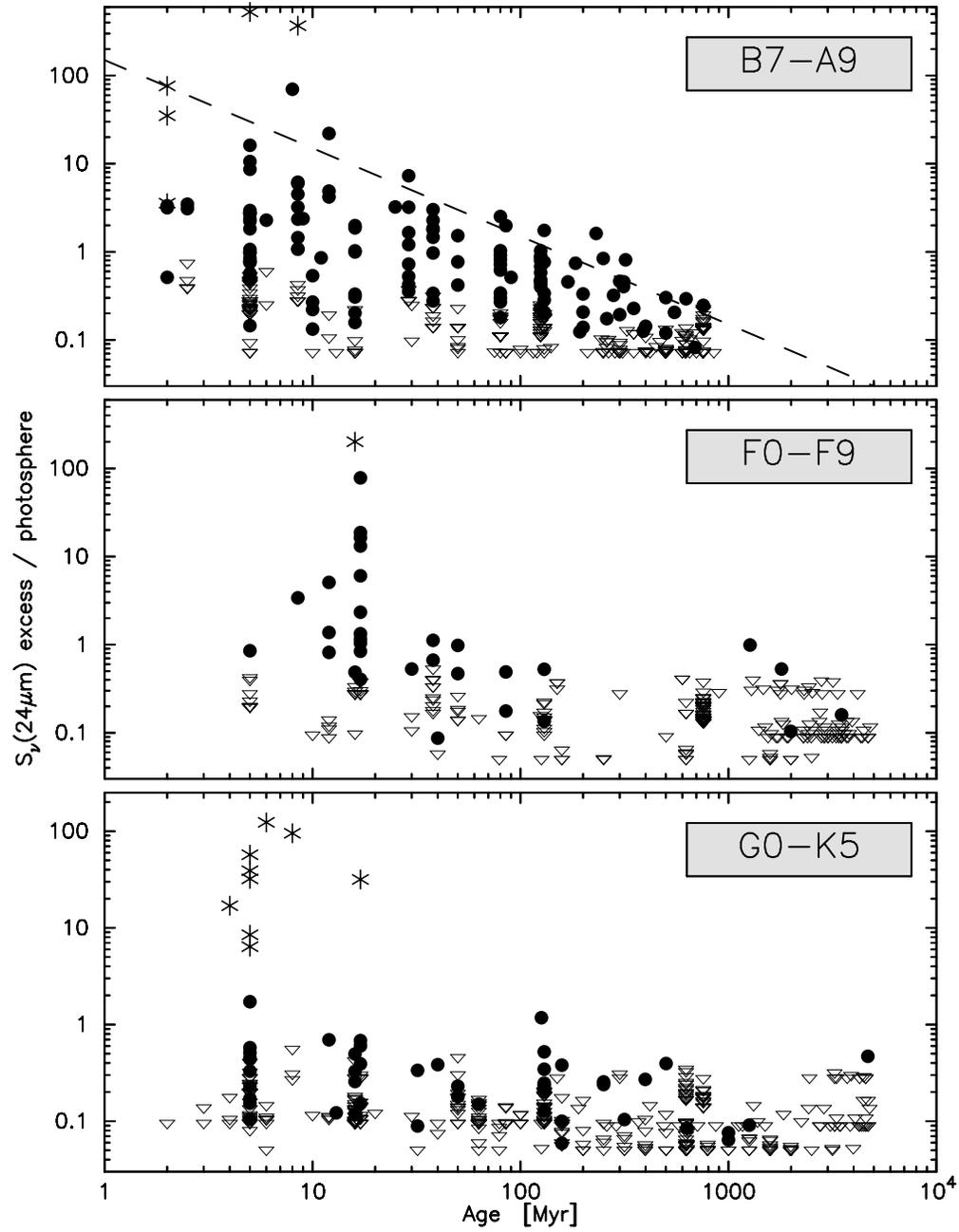}
\caption{
  \label{fig:excess24}
  Ratio of the 24\micron\ excess normalized by the photospheric 24\micron\ 
  flux density versus stellar age compiled from the regions summarized in 
  Table~\ref{tbl:lit}. Filled circles represent stars with a detected 
  24\micron\ excess that are likely debris disks, and open triangles 
  represent 3$\sigma$ upper limits for stars without excesses. Asterisks
  indicate optically thick disks (including ``anemic'' and ``evolved'' disks).
  Known Be stars and sources with contaminated 24\micron\ photometry have
  been omitted. The dashed line shows the upper envelope
  to the 24\micron\ excess from \citet{Rieke05} for ages older than
  5~Myr, and extrapolated here to younger ages.
}
\end{center}
\end{figure}

\begin{figure}
\begin{center}
\includegraphics[angle=-90,scale=0.8]{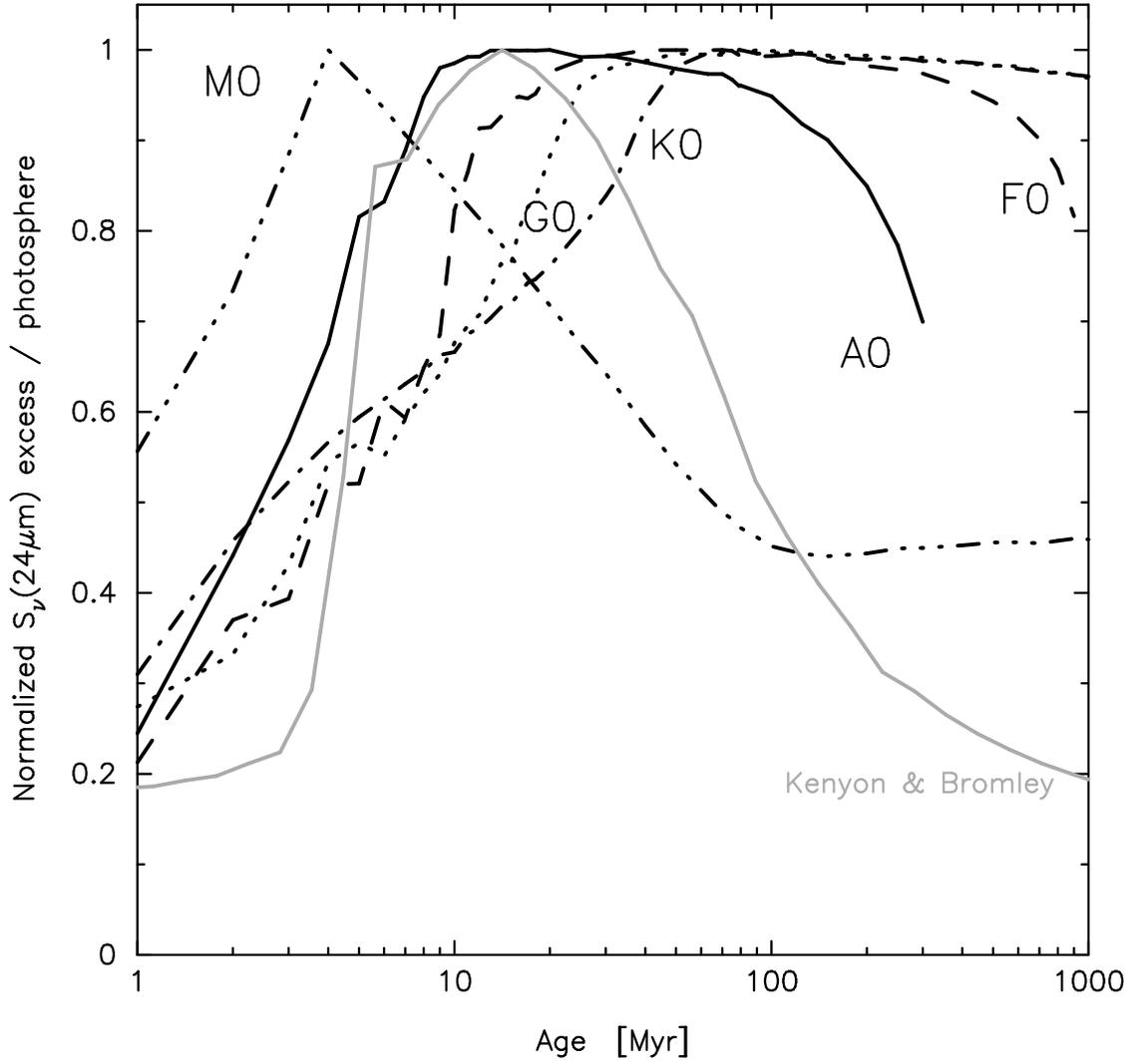}
\caption{
  \label{fig:sens}
  Expected temporal variation in the 24\micron\ excess relative to the stellar 
  photosphere for A0 (solid black curve), F0 (dashed), G0 (dotted), 
  K0 (dash-dot), and M0 stars (dash-dot-dot-dot) stars. The 24\micron\ excess 
  for a given curve has been normalized by the peak excess. The stellar 
  mass and luminosity will vary with age for a fixed spectral type as the 
  star evolves toward the main-sequence. The stellar parameters for a given age 
  and spectral type were estimated from the \citet{Siess00} pre-main-sequence 
  models. The model planetesimal belt is at an orbital radius of 15~AU. The 
  debris emission varies in time as stars evolve, which changes the stellar 
  heating and the radiation blowout size. These results show that if the mass 
  of the planetesimal belt is fixed, A-K stars younger than 10~Myr will have 
  a lower 24\micron\ excess since the smaller grains have been blown out of 
  the system by the greater stellar luminosity. For comparison,
  the solid gray curve shows the variation in the normalized 24\micron\ excess
  from a planetesimal belt with a mass of the minimum mass solar nebula that 
  extends between 30 and 150~AU around a 2\msun\ star with constant stellar 
  luminosity \citep{Kenyon08}.
}
\end{center}
\end{figure}

\begin{figure}
\begin{center}
\includegraphics[angle=-90,scale=0.8]{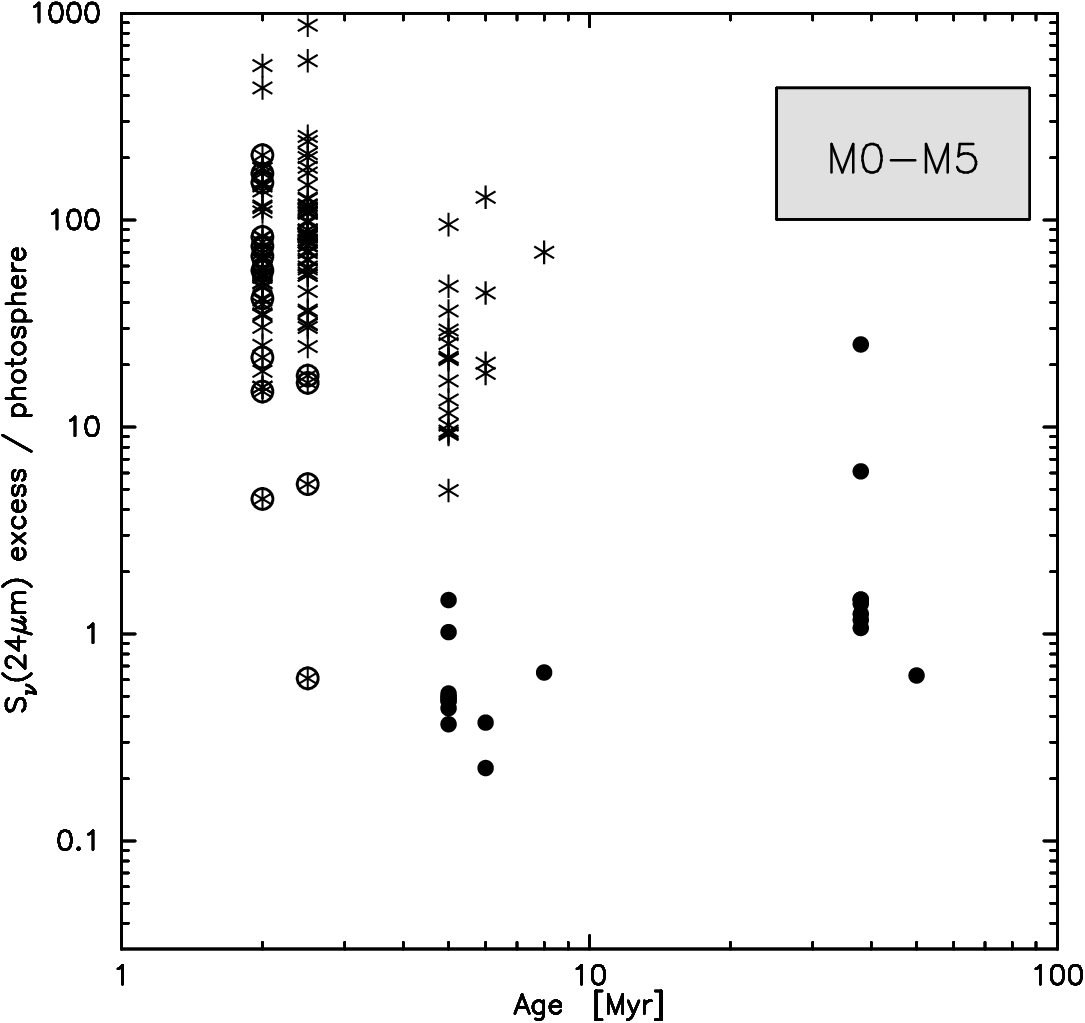}
\caption{
  \label{fig:mstars} 
  The 24\micron\ excess for M0-M5 stars as a function of stellar age.
  Excess sources are shown for IC~348 (age 2~Myr; \citealt{Lada06}),
  $\sigma$~Ori (2.5~Myr; \citealt{Hernandez07a},
  Upper Sco (5~Myr), 
  $\eta$~Cha (6~Myr; \citealt{Gautier08}),
  TW~Hydra moving group (8~Myr; \citealt{Low05}),
  NGC~2547 (38~Myr; \citealt{Forbrich08}),
  and 
  IC~2391 (50~Myr; \citealt{Siegler07}). Asterisks indicate stars where
  the 24\micron\ excess is thought to originate from an optically thick
  disk; circled asterisks are optically thick disks further subdivided
  as ``anemic'' or ``evolved'' disks. Filled circles indicate stars 
  classified as debris systems. M-stars with upper limits at 24\micron\ have
  been omitted for clarity. Not all surveys shown here are sensitive to the
  stellar photosphere, and this diagram should not be used to infer 
  the evolution of debris emission around M-type stars.
}
\end{center}
\end{figure}

Most \Spitzer\ studies have identified sources with excesses based on the 
presence of unusually red infrared colors (e.g. [K]-[24], [8]-[24]). 
Each study has a minimum detectable excess that depends on the depth of
the survey, the presence of nebulosity in sky background, the distance to the 
stellar sample, and the intrinsic brightness of the star. In nearly all cases, 
we adopted the limiting thresholds adopted in the original study to identify 
24\micron\ excesses, which is normally taken to be three or four times
the RMS scatter around the main locus of data points in a color-color diagram
or a stellar color that is thought to represent photospheric colors
(see, e.g., Figure~\ref{fig:ccd}). We then identified the latest spectral type
in which the internal uncertainty in the measured photospheric flux was less
than this RMS scatter. This criteria removed sources where the photometric
uncertainty was comparable to the excess being measured. If the threshold 
to identify an excess ($\equiv\Delta$) was expressed in magnitudes, the adopted 
threshold was estimated as $10^{\Delta/2.5}$.

The ages for clusters and associations were adopted from the literature based 
on a number of techniques, including pre-main-sequence evolutionary tracks,
lithium depletion, and the main sequence turnoff point. The age uncertainties
were adopted from the literature, but if no uncertainty was quoted, we 
assumed an age uncertainty of 0.3~dex in the logarithm of the age.
The age uncertainties for the clusters and associations do not reflect any
systematic uncertainty associated with the particular method. The ages for 
field stars were adopted from the studies presenting the \Spitzer\ data. The 
references for the ages are provided in Table~\ref{tbl:lit}.

A few studies are not included in this comparison since the observations could
not be analyzed in a consistent manner. First, we did not includes surveys
where the observations were not sensitive to the photosphere for spectral types
A or later, either because the regions are more distant than \about 900~pc
(Trumpler 37 and NGC 7160, \citealt{Sicilia06}; h \& $\chi$ Persei,
\citealt{Currie08a}; NGC~2262, \citealt{Currie08b}), or the observations were
exclusively of late spectral types \citep[$\lambda$ Orionis,][]{Barrado07}.
MIPS 24\micron\ observations of the $\gamma$ Velorum cluster were not
incorporated; while the observations were sensitive to the photosphere for
stars early than \about F8, the contamination from field stars is
non-negligible \citep{Hernandez08}. We also did not analyze MIPS observations
of weak-lined T Tauri stars since the samples excluded optically thick disks
\citep{Padgett06,Cieza07}.

Given the heterogeneous nature of the analyzes to identify \Spitzer\ excesses,
we occasionally had to re-analyze data from the original study to provide a
uniform analysis. The following discussion summarizes the regions
where we re-interpreted the observations.

{\it $\sigma$ Orionis}: From the photometry presented by \citet{Hernandez07a},
97\% (33 out of 34) stars brighter than $J$=10.2~mag were detected at 
24\micron, while only 4 of 7 stars between $J$=10.2 and 10.5~mag were 
detected. The completeness limit was then defined as $J$=10.2. For
stars fainter than $J$=10.2, the bluest star has a dereddened color of
$J-H=0.15$. Therefore, we assumed the survey is complete for $J-H<=0.15$,
corresponding to spectral type earlier than F0. 

{\it IC~348}: Of the eight A-type stars in IC~348 studied by \citet{Lada06},
six were detected by at 24\micron, one was not observed at 24\micron, and one
has an upper limit from locally bright background. Only 57\% of the FG stars
and less than 33\% of the KM stars in IC~348 have 24\micron\ measurements, and
therefore only the A-stars in IC~348 were included. We adopted a detection
threshold of 30\% for a 24\micron\ excess based on Figure~7 in \citet{Lada06}.

{\it NGC~2232}: \citet{Currie08b} estimated a 5$\sigma$ and 10$\sigma$ 
detection limit of 10.5~mag and 9.75~mag for their 24\micron\ survey. They
also find that the number of sources detected at 24\micron\ falls rapidly
for sources fainter than [24]=10.5. We therefore adopted [24]=10~mag as
a limiting magnitude, and find that 99\% of stars brighter than $J$=10~mag
have a MIPS 24\micron\ detection brighter than this limit. For a distance
of 340~pc, an age of 25~Myr, and a color excess of $E(B-V)$=0.07, $J$=10
corresponds roughly to a A8 star, which we adopt as the completeness limit
of the survey. The RMS scatter in the K-[24] color for stars brighter than
the completeness limit and not containing an infrared excess (as 
defined in \citealt{Currie08b}) is 0.09~mag, which was adopted as the 
photometric uncertainty.

{\it Praesepe}: From Figure~6 in \citet{Gaspar09}, we find that the number
of stars with unphysical blue stellar colors increases for stars redder than
$r-K_{\rm s}$=1.5. We consider only stars bluer than this limit, which 
corresponds to stars earlier than G5.

\clearpage
\LongTables

\begin{deluxetable}{lrc@{\extracolsep{10pt}}rlcc}
\tabletypesize{\scriptsize}
\tablewidth{0pt}
\tablecaption{MIPS 24 and 70\micron\ photometery\label{tbl:phot}}
\tablehead{
  \colhead{Source}  &
  \multicolumn{2}{c}{MIPS 24\micron\tablenotemark{1}}  &
  \multicolumn{2}{c}{MIPS 70\micron\tablenotemark{2}}  &
  \colhead{AORKEY} &
  Flags \\
  \cline{2-3}
  \cline{4-5}
  \colhead{}    &
  $S_\nu$ (mJy) &
  $\sigma$ (mJy) &
  $S_\nu$ (mJy) &
  $\sigma$ (mJy)
}
\startdata
HD 142987                                &     7.89 &     0.07 &      5.2 &     11.5 & 13908224 &      \\
HD 147810                                &     9.67 &     0.09 &     -8.7 &     10.2 & 13906432 &      \\
HD 149598                                &     5.56 &     0.05 &     11.0 &     21.6 & 13909504 &      \\
HIP 76071                                &    10.79 &     0.10 &    -11.4 &      8.8 & 13902848 &      \\
HIP 76310                                &   165.40 &     1.50 &    373.1 &     11.3 & 13904896 &      \\
HIP 76633                                &     9.26 &     0.08 &     12.0 &      9.0 & 13907968 &      \\
HIP 77457                                &     8.63 &     0.08 &     14.7 &     12.2 & 13906688 &      \\
HIP 77545                                &     9.48 &     0.09 &     59.3 &     58.5 & 13899264 &      \\
\enddata
\tablenotetext{1}{MIPS 24\micron\ photometry measured using a flux calibration factor of 0.0447 MJy/sr.}
\tablenotetext{2}{MIPS 70\micron\ photometry measured using a flux calibration factor of 702 MJy/sr.}
\tablenotetext{3}{Source names with the prefix [PZ99] and [PBB2002] are from \citet{Preibisch99} and \citet{Preibisch02}, respectively.}
\tablenotetext{4}{24\micron\ photometry is unreliable due to bright, extended nebulosity.}
\tablenotetext{5}{24\micron\ photometry is unreliable due to nearby star.}
\tablenotetext{6}{Source is extended at 24\micron}
\tablenotetext{7}{70\micron\ photometry is unreliable due to contaminating source or nebulosity.}
\tablenotetext{8}{Contaminating source(s) subtracted before 70\micron\ photometry was measured.}
\tablenotetext{9}{70\micron\ photometry measured on AOR 13903104.}
\end{deluxetable}

\clearpage

\begin{deluxetable}{llrrrrr@{\extracolsep{10pt}}ccl}
\tabletypesize{\scriptsize}
\tablewidth{0pt}
\tablecaption{Normalized Flux Ratios\label{tbl:excess}}
\tablehead{
  \colhead{Source}  &
  \colhead{SpT}     &
  \colhead{Ref}     &
  \colhead{A$_{\rm V}$} &
  \multicolumn{3}{c}{Ratio\tablenotemark{1}} &
  \multicolumn{2}{c}{Excess?} &
  \colhead{Type}\\
  \cline{5-7}
  \cline{8-9}
  \colhead{}    &
  \colhead{}    &
  \colhead{SpT} &
  \colhead{}    &
  \colhead{24\micron}    &
  \colhead{70\micron}    &
  \colhead{$\sigma$(70\micron)}    &
  \colhead{24\micron} &
  \colhead{70\micron}
}
\startdata
HD 142361                                & G3V          & 2   & 0.2    &     1.02 &      8.7 &     13.9 &   &   &  \\
HD 142987                                & G4           & 4   & 1.7    &     0.99 &      6.2 &     13.6 &   &   &  \\
HD 146516                                & G0IV         & 8   & 0.6    &     0.99 &      6.5 &     18.3 &   &   &  \\
HD 147810                                & G1           & 4   & 1.0    &     1.00 &     -8.3 &      9.8 &   &   &  \\
HD 149598                                & G0           & 4   & 0.7    &     1.13 &     18.5 &     36.2 &   &   &  \\
HIP 76071                                & B9V          & 2   & 0.3    &     0.98 &     -9.9 &      7.6 &   &   &  \\
HIP 76310                                & A0V          & 2   & 0.1    &    17.17 &    361.0 &     37.6 & Y & Y & debris\\
HIP 76633                                & B9V          & 2   & 0.3    &     1.30 &     12.1 &      9.1 &   &   &  \\
\enddata
\tablenotetext{1}{Ratio of the observed flux density to the photospheric value. The 24\micron\ ratios have an adopted uncertainty of 0.07.}
\tablerefs{
(1) \citet{Houk82};
(2) \citet{Houk88};
(3) \citet{Kunkel99};
(4) E. Mamajek, private communication;
(5) \citet{Preibisch98};
(6) \citet{Preibisch01};
(7) \citet{Preibisch02};
(8) \citet{Walter94}
}
\end{deluxetable}

\clearpage

\begin{deluxetable}{crrrr}
\tablewidth{0pt}
\tablecaption{Disk Types in Upper Sco\label{tbl:disktype}}
\tablehead{
  \colhead{Spt}  &
  \colhead{N(stars)}  &
  \multicolumn{3}{c}{N(disks)}\\
  \cline{3-5}
  \colhead{} &
  \colhead{} &
  \colhead{Be} &
  \colhead{Debris} &
  \colhead{Primordial}
}
\startdata
B     &  37  &  2  &  6  &  0\\
A     &  25  &  0  &  7  &  0\\
F     &   9  &  0  &  1  &  0\\
G     &  21  &  0  &  3  &  0\\
K     &  27  &  0  &  2  &  7\\
M     & 101  &  0  &  9  & 17\\
Total & 220  &  2  & 28  & 24
\enddata
\end{deluxetable}

\clearpage

\begin{deluxetable}{lccrcllc}
\tabletypesize{\scriptsize}
\tablewidth{0pt}
\tablecaption{Compilation of Debris Disk Surveys\label{tbl:lit}}
\tablehead{
  \colhead{Region}  &
  \colhead{Age}     &
  \colhead{Age} &
  \colhead{Nstars}  &
  \colhead{Minimum excess\tablenotemark{a}} &
  \multicolumn{2}{c}{SpT}     &
  \colhead{Photometry}\\
  \cline{6-7}
  \colhead{}        &
  \colhead{(Myr)}   &
  \colhead{Reference} &
  \colhead{}        &
  \colhead{(percent)} &
  \colhead{Early\tablenotemark{b}} &
  \colhead{Limit\tablenotemark{c}} &
  \colhead{Reference}
}
\startdata
CrA                         & 2-16           & 21    &   7 & 10 & G6   &  K5     & 4\\
IC 348                      &   2            & 17    &   6 & 30 & A0   &  A4     & 15     \\
$\sigma$ Ori                & 2.5            & 29    &  11 & 39 & B2   &  A3     & 13\\
Upper Sco                   &   5 $\pm$    1 & 25    & 220 & 32 & B0   &  M5     & 34\\
                            &                &       &   5 & 30 & F0   &  G5     & 5\\
                            &                &       &  16 & 10 & G0   &  K3     & 4\\
                            &                &       &   3 &  8 & B9   &  B8     & 32\\
Orion OB1b                  &   5 $\pm$    1 & 3     &  21 & 31 & B1   &  A8     & 12\\
\enddata
\tablenotetext{a}{The minimum percent 24\micron\ flux density above the photosphere used to identify an excess}
\tablenotetext{b}{Earliest spectral type in the survey}
\tablenotetext{c}{\ Latest spectral type where the minimum excess could be detected}
\tablenotetext{d}{\ We adopt an age of 50~Myr for IC~2391, and assume IC~2391 and IC~2602 have the same age following \citet{Stauffer97}.}
\tablerefs{
(1) \citet{Barrado04};
(2) \citet{Beichman06};
(3) \citet{Briceno05};
(4) \citet{Carpenter08};
(5) \citet{Chen05};
(6) \citet{Cieza08};
(7) \citet{Currie08b};
(8) \citet{Gaspar09};
(9) \citet{Gautier08};
(10) \citet{Gorlova06};
(11) \citet{Gorlova07};
(12) \citet{Hernandez06};
(13) \citet{Hernandez07a};
(14) \citet{King03};
(15) \citet{Lada06};
(16) \citet{Low05};
(17) \citet{Luhman03};
(18) \citet{Luhman04};
(19) \citet{Lyra06};
(20) \citet{Mamajek04};
(21) \citet{Meyer06};
(22) \citet{Naylor06};
(23) \citet{Odenkirchen98};
(24) \citet{Perryman98};
(25) \citet{Preibisch02};
(26) \citet{Preibisch08};
(27) \citet{Rebull08};
(28) \citet{Rieke05};
(29) \citet{Sherry08};
(30) \citet{Siegler07};
(31) \citet{Stauffer97};
(32) \citet{Su06};
(33) \citet{Trilling08};
(34) This study;
(35) \citet{vanRensbergen78};
(36) \citet{Zuckerman01};
(37) \citet{Zuckerman04}
}
\end{deluxetable}

\clearpage

\clearpage
\begin{center}
\includegraphics[scale=0.8]{figure2b.ps}
\clearpage
\includegraphics[scale=0.8]{figure2c.ps}
\end{center}

\end{document}